\documentclass[twocolumn, times, tighten, appendixfloats]{aastex7}

\usepackage{hyperref}
\usepackage{romannum}
\usepackage{amsmath}
\usepackage{float}
\usepackage{makecell}

\begin{document}

\pagenumbering{arabic}

\title{New Insight from the James Webb Space Telescope on Variable Active Galactic Nuclei}

%\title{Uncovering Varying Obscured AGN Candidates in the COSMOS Field with JWST/NIRCam Imaging}

% Variability study in the COSMOS field
% Variability in Galaxies as Seen by the James Webb Space Telescope
% Variable Galaxies: Type 1 or type 2 AGNs
% 
% Are there variable type 2 active galactic nuclei? 

\author[0000-0001-7957-6202]{Bangzheng Sun} 
\affiliation{Department of Physics and Astronomy, University of Missouri - Columbia, 701 S College Avenue, Columbia, MO 65201, USA}
\email{bangzheng.sun@mail.missouri.edu}

\author[0000-0001-7592-7714]{Haojing Yan}
\affiliation{Department of Physics and Astronomy, University of Missouri - Columbia, 701 S College Avenue, Columbia, MO 65201, USA}
\email{yanha@missouri.edu}

\author[0000-0003-0747-1780]{Wei Leong Tee} 
\affiliation{Steward Observatory, University of Arizona, 933 N. Cherry Ave, Tucson, AZ 85721, USA}
\email{wltee@arizona.edu}

\author[0000-0002-7633-431X]{Feige Wang}
%\affiliation{Steward Observatory, University of Arizona, 933 N. Cherry Ave, Tucson, AZ 85721, USA}
\affiliation{Department of Astronomy, University of Michigan, 1085 S. University Ave., Ann Arbor, MI 48109, USA}
\email{fgwang@umich.edu}

\begin{abstract}

   Variability detected in galaxies is usually attributed to their active 
galactic nuclei (AGNs). While all AGNs are intrinsically variable, the 
AGN unification model predicts that type~2 AGNs rarely vary because 
their engines are blocked by dust tori. Previous UV-to-near-IR 
variability studies largely support this expectation. Here, we present a 
variability study by James Webb Space Telescope (JWST) that reveals a more 
subtle picture. Using NIRCam imaging data from three surveys over 
$\sim$140~arcmin$^2$ in the COSMOS field, we found 117 galaxies with 
$\geq 4$$\sigma$ variability in the F356W band across a $\sim$2-year 
baseline. Cross-matching with the existing JWST spectroscopic data, 
we identified five of them at $z=0.19$--3.69 
(F356W corresponding to rest-frame $\lambda\approx0.76-2.97$~$\mu$m), 
which were all coincidentally 
observed by a NIRSpec program almost contemporaneously with the last imaging 
epoch. One additional variable was identified at $z=0.90$ using the archival 
Keck telescope data. These six objects form our spectroscopic subsample.
%{\bf whose variability in F356W corresponds to rest-frame 
%$\lambda\approx0.76-2.97\mu$m.}
Interestingly, two reside in close-pair environments, while two others
form a close pair themselves. 
Most of their light curves can hardly be explained by nuclear transients, 
and AGN variability is a more plausible cause. However, among these six 
objects, (1) only one shows broad Bracket and Pfund series permitted 
lines ($\Delta v > 1000$~km~s$^{-1}$) indicative of a type~1 AGN; 
(2) two show narrow permitted lines (H$\alpha$ and/or 
He~I$\lambda10830$) consistent with type~2 AGNs, with another one likely 
type~2 based on the host galaxy properties; and (3) two others, which form a 
pair, show no emission lines. Our results add more challenges to the 
unification model.

\end{abstract}

%\keywords{...}

\section{Introduction} \label{sec:intro}

    It is well known that active galactic nuclei (AGNs) could stochastically
vary in brightness at all wavelengths on any time scales
\citep[see e.g.,][]{Ulrich1997}. The AGN variability mechanisms, while still
under debate, are widely believed to be associated with the accretion disk
\citep[e.g.,][]{Rees1984, Kawaguchi1998, Hawkins2002, Wilhite2008, Kelly2009, Dexter_Agol2011}. In the canonical AGN unification model
\citep[][]{Antonucci1993,Urry1995,Netzer2015}, 
type~1 (unobscured) AGNs could vary in both continuum and broad permitted 
emission lines because the accretion disk and the broad line region (BLR) are 
exposed to us directly. On the contrary, type~2 (obscured) AGNs are expected to 
show minimal or no variability in continuum and narrow lines because their 
accretion disk and BLR are blocked from view by the dust torus and the narrow 
line region (NLR) is relatively stable due to the long recombination time scales.
Such expectations have been largely confirmed by observations 
\citep[e.g., ][]{Yip2009}. 

     Nevertheless, several variability surveys over the past decade have found
that a small fraction of type~2 AGNs do vary on multi-year timescales. 
For example, \citet[][]{Choi2014} found detectable long-term optical variability 
in two of their six spectroscopically identified type~2 quasars, 
while \citet[][]{Cartier2015} reported that $\sim21\%$ of narrow-line AGNs 
in the QUEST-La Silla survey showed variability over a six-year baseline. 
%Near-infrared monitoring further shows that the variability of Type II AGNs 
%would decline at longer wavelengths because dust-reprocessed emission 
%from the dust torus attenuates the intrinsic variability \citep[][]{Sanchez2017}. 
Recently, \citet[][]{LopezNavas2023} used the Zwicky Transient Facility data over
2.5 years to investigate the variability of $\sim$15,000 AGNs from the Sloan
Digital Sky Survey and found that $\sim$11\% type~2 AGNs have optical variations.
\citet[][]{Kovacevic2025} used the time-series data from TESS and ASAS-SN and 
showed that while the majority of Seyfert 2 galaxies do not vary, a very small 
number of them do undergo significant optical variation. Such variable type~2 
sources pose a challenge to the unification model and could potentially lead to 
the identification of ``true'' type~2 AGNs, whose lack of BLRs is genuine but 
not due to the orientation effect \citep[][]{Tran2003}. However, it has also 
been argued that many variable type 2 sources might simply be misclassified 
type~1 \citep[e.g.,][]{Barth2014, Sanchez2017, ZhangXG2023, LopezNavas2023} and 
therefore do not violate the unification picture. On the other hand, the newly 
established changing-look AGN class 
\citep[CLAGN; see][for a recent review]{Ricci2023}, 
which directly confronts the unification model, leads to a new view of
variable type~2 AGNs: they could be CLAGNs, and their variations occur when they
``turn on'' or ``turn off'' their BLRs \citep[e.g.,][]{LopezNavas2023}. For this
reason, variability search among type~2 AGNs could be an effective method of 
selecting CLAGNs \citep[e.g.,][]{LopezNavas2022,WangShu2024}.

   The aforementioned AGN variability studies are mostly on bright AGNs that 
are luminous and/or nearby. There have been a few searches for variable AGNs 
at the low luminosity end using the Hubble Space Telescope
\citep[][]{Sarajedini2003, Cohen2006, Klesman2007, Villforth2010, Pouliasis2019, OBrien2024},
however, spectroscopic confirmations on the candidate samples are scarce due to
the faintness of the targets. Thanks to its superb sensitivity and spatial 
resolution, the James Webb Space Telescope (JWST) will push the AGN variability 
study to an unprecedented limit in both imaging and spectroscopy. After $\sim$3 
years of operation, its imaging surveys by the NIRCam instrument have built a 
sufficient time baseline for long-term variability search. In the meantime, its 
spectroscopy programs have also accumulated a large number of spectroscopic 
samples for cross identification. Recently, \citet[][]{DeCoursey2025} used such 
data and found a variable galaxy at $z=5.274$ dimming from 26.05 to 26.24~mag
over a year. While it is still uncertain whether the variability was due to a 
transient event (e.g., a supernova) or an AGN, the discovery confirms the power 
of JWST in studying variability at high redshifts and in the low luminosity 
regime.

    To this end, we have carried out a search for variables using the JWST data 
in the COSMOS field, and this paper reports our initial findings. The structure 
of the paper is as follows.
We describe the observations and the data in Section 2. The search for variable
objects and the spectroscopic subsample are given in Section 3. The results are
discussed in Section 4, and we conclude in Section 5. All magnitudes quoted in 
this work are in the AB system \citep[][]{Oke1983_ABmag}, and all coordinates 
are in the ICRS frame and Equinox 2000. We adopted a flat $\Lambda$CDM cosmology
with $\rm H_0=71~km~s^{-1}~Mpc^{-1}$, $\rm \Omega_{m}=0.27$, and 
$\rm \Omega_{\Lambda}=0.73$.

\section{JWST Data Description}

%A major objective of this work is to study the discovered variable objects
%using spectroscopic data. 
   As one of the most well-studied areas for extragalactic sciences, the COSMOS
field has been targeted by various JWST programs since Cycle 1. Three large
NIRCam imaging programs established a baseline over two years, which enables 
this study. In addition, the field also has sufficient spectroscopic data that 
allow us to assemble a subsample to further investigate the nature of the 
variable galaxies. The relevant data are described below.

\subsection{Imaging Data}

   The NIRCam imaging data used in this work were obtained between 2022 December
26 and 2025 April 22 by three programs: 
(1) Public Release IMaging for Extragalactic Research (PRIMER; PID 1837; 
PI J. Dunlop), 
(2) COSMOS-Web \citep[PID 1727, PIs J. Kartaltepe \& C. Casey;][]{Casey2023},  
and (3) COSMOS-3D (PID 5893; PI K. Kakiichi). 
The PRIMER program covered $\sim$140~arcmin$^2$, which is the smallest among the
three; therefore, we limited our search to this area. 

$\bullet$ PRIMER-COSMOS (PRM): our reference epoch. PRIMER utilized eight 
NIRCam filters, which are F090W, F115W, F150W, and F200W in the short wavelength 
(SW) channel and F277W, F356W, F410M, and F444W in the long wavelength (LW) 
channel. The NIRCam imaging of this program was done in two periods of time 
approximately $\sim$4 months apart and covered two different sections of the 
field. We therefore divided the entire program into two separate epochs, which 
we refer to as PRM.1 (Observation 4) and PRM.2 (Observation 3). The observation 
start times were 2022 December 26 and 2023 April 27, respectively. 

$\bullet$ COSMOS-3D (C3D): our discovery epoch. Most of the COSMOS-3D data 
were obtained approximately 2 years after PRIMER. Because the NIRCam imaging in 
this program was done as part of its main NIRCam wide-field slitless spectroscopy
(WFSS) observations, only three filters were utilized: F115W, F200W and F356W. 
We also divided the data from this program into two separate epochs based on 
their overlaps with the two PRM epochs, which we refer to as C3D.1 (Observation 
18) and C3D.2 (Observations 10, 11, 14, 15). The two epochs started 
on 2024 December 21 and 2025 April 15, respectively, and therefore are 726 and 
719 days apart from PRM.1 and PRM.2, respectively. 

$\bullet$ COSMOS-Web (WEB): an intermediate epoch. COSMOS-Web supplies one 
or two additional time samplings in four passbands: F115W, F150W, F277W, and 
F444W. While these data are shallower than PRIMER and COSMOS-3D in most cases, 
they still provide useful constraints on the light curves. Again, we divided the 
data into three epochs based on the coverage and observation times: 
WEB.1 (Observations 14, 16, 18, 20, 51--61), 
WEB.2 (Observations 91, 93, 95, 97), and WEB.3 (Observation 159). We note that 
observation 159 (WEB.3) was a re-observation of Observation 18 (part of WEB.1), 
and thus only two filters were utilized instead of four. 
The observations started on 2023 April 15, 2023 December 28, and 2024 April 17 
for the three epochs, respectively. 

   Table~\ref{tab:obs} summarizes all these data, including the filters used 
and the observation times. 

   We reduced these data on our own using the JWST pipeline 
\citep[][]{Bushouse24_jwppl} version 1.18.0 in the calibration ``context'' of 
{\tt jwst\_1364.pmap}, following the procedures outlined in 
\citet[][]{Yan_highz_23}. We adopted the Gaia Data Release 3 for astrometry, 
and the astrometric alignment precision is $<15$~mas for all individual exposures. 
The final images have the pixel scale of 60~mas and are all aligned in the pixel 
coordinates. For each science image, we also derived the ``root mean square'' 
(RMS) map from the weight image using {\sc astroRMS}
\footnote{Courtesy of M. Mechtley; \url{https://github.com/mmechtley/astroRMS}},
which accounts for the correlated noise due to pixel resampling. These RMS maps
were used for photometric error derivation.

%{\color{red} special treatments to the two MSA spectra with two sources in the same slit setup}

%in particular these four that used the MSA mode:
%PIDs 1214 (PI N. Luetzgendorf), 2565 (PI K. Glazebrook), 6368 (CAPERS; 
%PI M. Dickinson), and 6585 (PI D. Coulter). We searched the data from these 
%programs and found matches in PIDs 1214 and 

\subsection{Spectroscopic Data}\label{sec:nirspec_data} 

   The COSMOS field has been observed by a few JWST NIRSpec programs, and we 
have reduced all their data. As described in the next section, we found matches
to our variables in the ongoing CANDELS-Area Prism Epoch of Reionization Survey 
(CAPERS; PID 6368; PI. M. Dickinson). This program uses the NIRSpec 
micro-shutter assembly (MSA) for multi-object spectroscopy under the PRISM/CLEAR
disperser/filter setup (hereafter the `PRISM'' mode), which has the spectral 
resolution ($R$) ranging from $\sim$30 to 300 over the entire NIRSpec wavelength 
coverage of 0.6--5.3~$\mu$m. The targets are covered by three-shutter slitlets, 
and the observations are done in a three-point nodding pattern. The existing 
data were all taken in 2025 April and May, which were almost contemporaneous 
with the C3D.2 epoch and were only $\sim$4--5 months after the C3D.1 epoch. 

    We reduced these data following the procedures described in 
\citet[][]{Yan2024_dz5289}. Briefly, we first processed the data using the JWST 
pipeline through the {\tt calwebb\_detector1} step, and then 
the output ``rate.fits'' files were further reduced using the {\sc msaexp} 
package \citep[version 0.9.2;][]{Brammar23_msaexp}, which provides an end-to-end 
reduction till the final spectra extraction. There were a few special cases 
where we had to manually adjust the background estimate and/or the extraction 
procedures, which will be described in more details when we discuss the 
individual objects in Section 3.2.

    As a side note, we also reduced the COSMOS-3D NIRCam WFSS data. However, 
none of our variables presented in the next section have high S/N 
emission lines detected in these data, presumably due to the limited wavelength 
coverage. %We will not discuss these data here.

%\subsection{NIRCam WFSS Spectra}\label{sec:wfss_spec}
%The PRIMER-COSMOS field was observed by the NIRCam instrument
%in the wide-field slitless spectroscopy (WFSS) mode
%as the main observation of the COSMOS-3D program.
%These observations utilized the Grism R setting,
%in which the light is dispersed along the direction of the detector rows
%in the LW channel.
%It has a resolution of $R\approx1600$ at $\sim4\mu{\rm m}$.
%
%The observations were carried out simultaneously with their NIRCam imaging
%using the F444W filter.
%To reduce these data, we retrieved the Level 1b data
%from the Mikulski Archive for Space Telescopes (MAST)
%and ran them through the {\tt calwebb\_detector1} step
%of the JWST data reduction pipeline (version 1.18.0 in the context of jwst\_1364.pmap) and obtained the ``rate.fits" files.
%Then, we followed the procedures by \citet[][]{Sun2023}
%to further process the data and extract the spectra.
%All single exposures were registered to GAIA DR3
%so that the astrometry is consistent with the imaging data.

%\subsection{Other Spectroscopic Identifications}

\begin{table*}[hbt!]
    \raggedright
    \caption{Summary of NIRCam observations in each epoch}
    \begin{tabular}{cccccc} \hline
        Epoch & \makecell{NIRCam Passbands} & Obs. Start (UT) & Obs. End (UT) \\ \hline 
        PRM.1 & \makecell{F090W,F115W,F150W,F200W\\F277W,F356W,F410M,F444W} & 2022-12-26 18:49 & 2023-01-06 04:50 \\ \hline 
        WEB.1 & \makecell{F115W,F150W,F277W,F444W} & 2023-04-15 06:04 & 2023-04-15 10:24 \\ \hline 
        PRM.2 & \makecell{F090W,F115W,F150W,F200W\\F277W,F356W,F410M,F444W} & 2023-04-27 02:56 & 2023-05-26 19:49 \\ \hline 
        WEB.2 & \makecell{F115W,F150W,F277W,F444W} & 2023-12-28 05:22 & 2024-01-06 19:22 \\ \hline
        WEB.3 & \makecell{F150W,F444W} & 2024-04-17 12:46 & 2024-04-17 14:09 \\ \hline 
        C3D.1 & \makecell{F115W,F200W,F356W} & 2024-12-21 08:52 & 2024-12-21 22:59 \\ \hline 
        C3D.2 & \makecell{F115W,F200W,F356W} & 2025-04-15 21:03 & 2025-04-22 06:11 \\ \hline 
    \end{tabular}
    \tablecomments{
    ``Obs. Start" and ``Obs. End" are the times (in UT) of the beginning of the 
    first observation and the end of the last observation for each epoch,
    respectively.
    }
    \label{tab:obs}
\end{table*}

\section{Variable Objects and Spectroscopic Identification}

\subsection{Variable Search}\label{sec:var_search}

   Our initial search for variables was carried out by comparing the photometry
on the images from PRM.1/PRM.2 and C3D.1/C3D.2, which form two pairs in terms of
sky coverage: C3D.1 covers the PRM.1 area, while C3D.2 covers the PRM.2 area. 
The two epochs in these two pairs were both separated by $\sim$2 years. We will
refer to these two areas as ``C3DvarN'' and ``C3DvarS'', respectively, where 
``N'' (north) and ``S'' (south) stand for their relative locations in the sky to 
each other. 

   The photometry was done by running {\sc SExtractor} 
\citep[][]{Bertin1996} in the dual-image mode, and the F356W images in the C3D 
epochs were used as the detection images. While there are two other common 
bands (F115W and F200W) between these two data sets, we chose the F356W band 
because (1) the background of the F356W images is notably cleaner than the two 
SW bands and (2) the F356W images are the deepest among the three bands. 
To add more constraints on the light curves, the photometry was also done on
the COSMOS-Web images in the same manner.

\begin{figure}
    \centering
    \includegraphics[width=\linewidth]{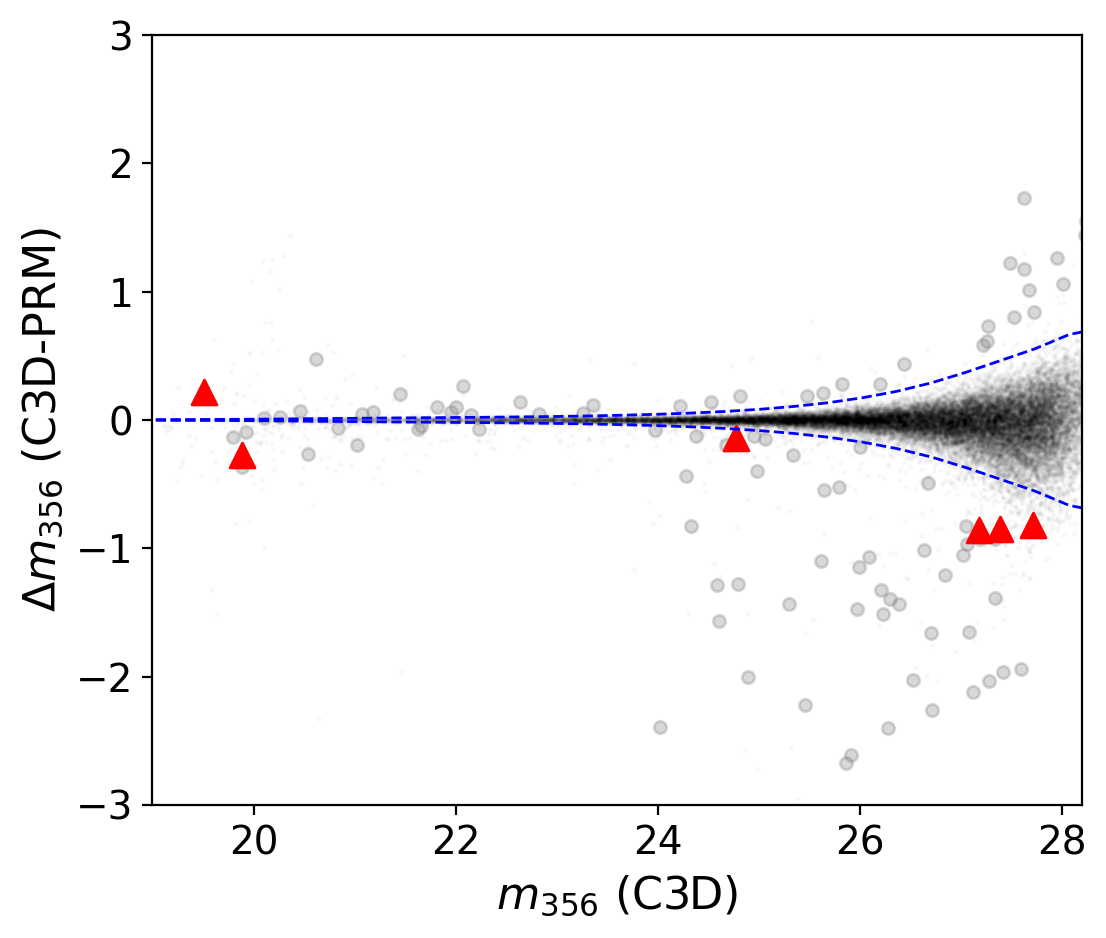}
    \caption{F356W magnitude differences between the C3D and PRM epochs versus
    C3D F356W magnitude. The small back dots are all sources in the field. The
    blue dashed curves represent the 4$\sigma$ statistical boundary of the 
    distribution. The large gray, filled circles indicate the 117 objects 
    retained in the variable sample. There are more brightening variables than
    fading ones because the C3D images were used for detections. The six
    variables that form the spectroscopic sample are shown as the red triangles. 
    }
    \label{fig:mag_dmag}
\end{figure}

   Variable objects were selected based on the variation of the F356W flux
between the PRM and C3D epochs. We adopted \texttt{FLUX\_ISO}, which is the 
flux measured in the isophotal aperture. A source was flagged as a candidate 
variable if the flux difference is at least 4 times the total measurement 
uncertainty, i.e., 
$|f_{\rm C3D}-f_{\rm PRM}|\geq4\times\sqrt{\sigma_{C3D}^2+\sigma_{PRM}^2}$.
Figure~\ref{fig:mag_dmag} shows the distribution of the F356W magnitude 
differences for all sources between the PRM and C3D epochs as a function of the 
C3D F356W magnitudes. For illustration, the 4~$\sigma$ limits of the 
distribution 
(calculated in the flux density domain and converted to the magnitude domain)
are represented by the dashed blue curves. Note that an object being flagged
or not was based on the total measurement uncertainty of this particular source
(see the equation above), which is not exactly the same as given by the blue 
curves that indicate the statistical boundary based on all sources.
Our initial selection yielded $\sim$1100 candidate variables, which were then 
visually inspected in the per-epoch images as well as in the difference images
between epochs. In total, 117 were retained, which are shown in 
Figure~\ref{fig:mag_dmag} as the gray, filled circles.
The rejected ones are mostly fake sources around bright objects (especially 
stars) and spurious detections due to cosmic-ray residuals, hot pixels and 
other cosmetic defects.
Among the retained objects, 13 are transients that appeared in only one
epoch. 

\begin{table*}[hbt!]
    \raggedright
    \scriptsize
    \caption{Observation details of the spectroscopic sample}
    \begin{tabular}{lccccccc} \hline
        SID & $z_{\rm spec}$ & R.A. (deg) & Decl. (deg) & Epoch & First Obs. (UT) & Last Obs. (UT) & 5$\sigma$ F356W Depth \\ \hline 
        C3DvarN\_0173 & $3.69\pm0.01$ & 150.1496839 & 2.4212437 & PRM.1 & 2023-01-02 13:58 & 2023-01-06 04:50 & 28.87 \\ 
        & & & & WEB.2 & 2024-01-06 13:18 & 2024-01-06 14:35 & ... \\ 
        & & & & WEB.3 & 2024-04-17 12:46 & 2024-04-17 14:09 & ... \\ 
        & & & & C3D.1 & 2024-12-21 18:39 & 2024-12-21 21:40 & 28.51 \\ 
        \hline 
        
        C3DvarS\_0007 & $2.09\pm0.02$ & 150.1310565 & 2.1667905 & WEB.1 & 2023-04-15 09:12 & 2023-04-15 10:24 & ... \\ 
        & & & & PRM.2 & 2023-04-27 12:39 & 2023-04-27 15:00 & 28.04 \\ 
        & & & & C3D.2 & 2025-04-21 21:11 & 2025-04-22 06:11 & 28.33 \\ 
        \hline 
        %C3DvarS\_0187 & $1.83\pm0.01$ & 150.1332270 & 2.2254686 & PRM.2 & 2023-04-27 12:39 & 2023-05-22 14:34 & 28.86 \\ 
        %& & & & WEB.2 & 2023-12-28 11:18 & 2023-12-28 12:35 & ... \\
        %& & & & C3D.2 & 2025-04-21 21:11 & 2025-04-21 22:36 & 28.06 \\ 
        %\hline 
        C3DvarS\_0722 & $2.8^\triangle$ & 150.0973109 & 2.3679997 & PRM.1 & 2023-01-03 21:19 & 2023-01-03 22:53 & 28.07 \\ 
        & & & & PRM.2 & 2023-05-18 02:24 & 2023-05-18 10:29 & 28.79 \\ 
        & & & & WEB.2 & 2023-12-29 10:54 & 2023-12-29 12:08 & ... \\ 
        & & & & C3D.2 & 2025-04-15 22:38 & 2025-04-15 23:58 & 28.20 \\ 
        \hline 
        C3DvarS\_0723 & $2.8^\triangle$ & 150.0980403 & 2.3681775 & PRM.1 & 2023-01-03 21:19 & 2023-01-03 22:53 & 28.18 \\ 
        & & & & PRM.2 & 2023-05-18 02:24 & 2023-05-18 10:29 & 28.88 \\ 
        & & & & WEB.2 & 2023-12-29 10:54 & 2023-12-29 12:08 & ... \\ 
        & & & & C3D.2 & 2025-04-15 22:38 & 2025-04-15 23:58 & 28.20 \\ 
        \hline 
        C3DvarS\_0886 & $0.187\pm0.004$     & 150.1379773 & 2.2916546 & PRM.1 & 2023-01-03 19:43 & 2023-01-03 21:19 & 28.03 \\ 
        & & & & PRM.2 & 2023-04-27 15:00 & 2023-05-06 15:37 & 28.83 \\ 
        & & & & WEB.2 & 2023-12-28 11:18 & 2023-12-28 12:35 & ... \\ 
        & & & & C3D.2 & 2025-04-18 14:32 & 2025-04-18 16:13 & 28.10 \\ 
        \hline 
        C3DvarN\_0168 & $0.9^*$       & 150.1583496 & 2.4153917 & PRM.1 & 2022-12-26 22:14 & 2023-01-02 15:32 & 28.67 \\ 
        & & & & WEB.2 & 2024-01-06 13:18 & 2024-01-06 14:35 & ... \\ 
        & & & & WEB.3 & 2024-04-17 12:46 & 2024-04-17 14:09 & ... \\ 
        & & & & C3D.1 & 2024-12-21 18:39 & 2024-12-21 21:40 & 28.42 \\ 
        \hline 
    \end{tabular}
    \tablecomments{
    (1) The spectroscopic redshifts are mainly based on the JWST NIRSpec data. 
    The only exceptions is \texttt{C3DvarN\_0168} (marked with ``*''), whose 
    redshift is from the Keck 10K-DEIMOS survey. 
    The two objects flagged with ``$\triangle$'' (in the second column) have no
    emission lines present, and their redshifts are solely based on a steep
    absorption feature identified as the 4000\AA~break.
    (3) Times in ``First Obs." and ``Last Obs." are the start of the first 
    observation and the end of the last observation covering each target, 
    respectively. The 5~$\sigma$ F356W limits ($m_{\rm 356}$) were calculated
    using the RMS maps within a circular aperture of $r=0\farcs2$. 
    }
    \label{tab:var_info}
\end{table*}

\begin{table*}[hbt!]
    \raggedright
    \tiny
    \caption{Photometry of the six variables in the spectroscopic sample}
    \begin{tabular}{llcccccccc}
        SID & Epoch (Day) & $m_{\rm 090}$ & $m_{\rm 115}$ & $m_{\rm 150}$ & $m_{\rm 200}$ & $m_{\rm 277}$ & $m_{\rm 356}$ & $m_{\rm 410}$ & $m_{\rm 444}$ \\ \hline 
        C3DvarN\_0173 & $\lambda_{\rm rest}$ (\AA) & 1923 & 2461 & 3200 & 4249 & 5940 & 7597 & 8704 & 9426 \\ 
         & PRM.1 (0) & $25.20\pm0.04$ & $25.04\pm0.04$ & $25.19\pm0.04$ & $25.00\pm0.03$ & $24.50\pm0.01$ & $24.77\pm0.02$ & $24.71\pm0.03$ & $24.53\pm0.02$ \\ 
         & WEB.2 (367/78) & ... & $25.06\pm0.14$ & ... & ... & $24.52\pm0.02$ & ... & ... & $24.45\pm0.03$ \\ 
         & WEB.3 (469/100) & ... & ... & $24.90\pm0.07$ & ... & ... & ... & ... & $24.44\pm0.02$ \\ 
         & C3D.1 (717/153) & ... & $24.98\pm0.06$ & ... & $24.78\pm0.02$ & ... & $24.63\pm0.02$ & ... & ... \\ 
        \hline 
        C3DvarS\_0007 & $\lambda_{\rm rest}$ (\AA) & 2919 & 3735 & 4858 & 6450 & 9016 & 11531 & 13210 & 14307 \\
         & WEB.1 (0) & ... & $27.16\pm0.23$ & $27.56\pm0.26$ & ... & $28.00\pm0.21$ & ... & ... & $27.68\pm0.18$ \\ 
        & PRM.2 (12/4) & $>27.92$ & $27.54\pm0.41$ & $27.38\pm0.30$ & / & $27.83\pm0.24$ & $28.23\pm0.21$ & $27.83\pm0.38$ & $27.62\pm0.23$ \\ 
        & C3D.2 (726/235) & ... & $26.92\pm0.18$ & ... & $27.12\pm0.08$ & ... & $27.38\pm0.10$ & ... & ... \\ 
        \hline 
        C3DvarS\_0722 & $\lambda_{\rm rest}$ (\AA) & 2374 & 3037 & 3950 & 5245 & 7332 & 9376 & 10742 & 11634 \\
        & PRM.1 (0) & $>27.79$ & $>27.84$ & ... & ... & $28.70\pm0.32$ & $28.22\pm0.16$ & ... & ... \\ 
        & PRM.2 (135/36) & $>28.51$ & $29.01\pm0.65$ & $28.79\pm0.44$ & $29.51\pm0.70$ & $29.10\pm0.31$ & $28.53\pm0.14$ & $28.55\pm0.31$ & $28.70\pm0.24$ \\ 
        & WEB.2 (360/95) & ... & $>27.51$ & $>27.78$ & ... & $>28.73$ & ... & ... & $>28.53$ \\ 
        & C3D.2 (833/219) & ... & $>27.70$ & ... & $27.48\pm0.16$ & ... & $27.71\pm0.12$ & ... & ... \\ 
        \hline 
        C3DvarS\_0723 & $\lambda_{\rm rest}$ (\AA) & 2374 & 3037 & 3950 & 5245 & 7332 & 9376 & 10742 & 11634 \\
        & PRM.1 (0) & $>27.81$ & $>27.84$ & ... & ... & $28.25\pm0.24$ & $28.11\pm0.16$ & ... & ... \\ 
        & PRM.2 (135/36) & $>28.51$ & $>28.55$ & $27.76\pm0.19$ & $27.97\pm0.20$ & $28.39\pm0.16$ & $28.04\pm0.09$ & $27.91\pm0.18$ & $28.09\pm0.15$ \\ 
        & WEB.2 (360/95) & ... & $>27.06$ & $>27.41$ & ... & $28.08\pm0.27$ & ... & ... & $27.95\pm0.28$ \\ 
        & C3D.2 (833/219) & ... & $>27.92$ & ... & $27.41\pm0.16$ & ... & $27.18\pm0.08$ & ... & ... \\ 
        \hline 
        C3DvarS\_0886$^\dagger$ & $\lambda_{\rm rest}$ (\AA) & 7599 & 9722 & 12645 & 16790 & 23471 & 30017 & 34389 & 37245 \\
        & PRM.1 (0) & $20.72\pm0.01$ & $20.21\pm0.01$ & $19.86\pm0.01$ & $19.59\pm0.01$ & $19.88\pm0.01$ & $20.05\pm0.01$ & $19.98\pm0.01$ & $20.03\pm0.01$ \\ 
        & PRM.2 (118/99) & $20.73\pm0.01$ & $20.22\pm0.01$ & $19.88\pm0.01$ & $19.64\pm0.01$ & $19.96\pm0.01$ & $20.16\pm0.01$ & $20.15\pm0.01$ & $20.16\pm0.01$ \\ 
        & WEB.2 (359/302) & ... & $20.05\pm0.01$ & $19.71\pm0.01$ & ... & $19.58\pm0.01$ & ... & ... & $19.73\pm0.01$ \\ 
        & C3D.2 (836/704) & ... & $20.14\pm0.01$ & ... & $19.51\pm0.01$ & ... & $19.89\pm0.01$ & ... & ... \\ 
        \hline 
        C3DvarN\_0168 & $\lambda_{\rm rest}$ (\AA) & 4747 & 6074 & 7900 & 10489 & 14663 & 18753 & 21484 & 23268 \\
        & PRM.1 (0) & $22.58\pm0.02$ & $21.61\pm0.01$ & $20.97\pm0.01$ & $20.35\pm0.01$ & $19.75\pm0.01$ & $19.51\pm0.01$ & $19.41\pm0.01$ & $19.45\pm0.01$ \\ 
         & WEB.2 (373/196) & ... & $21.67\pm0.02$ & $21.10\pm0.01$ & ... & $19.91\pm0.01$ & ... & ... & $19.65\pm0.01$ \\ 
         & WEB.3 (475/250) & ... & ... & $20.92\pm0.01$ & ... & ... & ... & ... & $19.72\pm0.01$ \\ 
         & C3D.1 (723/381) & ... & $21.66\pm0.01$ & ... & $20.41\pm0.01$ & ... & $19.72\pm0.01$ & ... & ... \\ 
         \hline 
    \end{tabular}
    \tablecomments{
    (1) The first row for each variable shows the rest-frame wavelength 
    estimated using the effective wavelength of each band. 
    (2) Day 0 for each variable is defined as the date of the first observation; 
    the two numbers in the parentheses are the observed and rest-frame days since Day 0. 
    (3) The missing photometry is labeled by ``..." if the image in this band 
    does not cover the target, or by ``/" if the image is severely contaminated 
    by bad pixels, cosmic ray residuals, etc. 
    (4) For a non-detection (S/N~$<2$), the $2\sigma$ upper limit measured 
    within a circular aperture of $r=0\farcs2$ is quoted instead. 
    (5) For \texttt{C3DvarS\_0886}, the photometry reported here were done at
    the center within a circular aperture of $r=0\farcs24$ to better reflect
    the variation in its nuclear region.
    }
    \label{tab:phot}
\end{table*}

\subsection{Spectroscopic Sample} \label{sec:spec_id}

   We searched the publicly available JWST NIRSpec data in the Mikulski Archive
for Space Telescopes (MAST) for matches to our variables. Five matches were 
found, which are all from the CAPERS program (see
Section~\ref{sec:nirspec_data}). We also searched the data from the 10K-DEIMOS 
survey done at the Keck telescope \citep[][]{Hasinger2018_10k-deimos} and found 
one more match. These six variables form our spectroscopic sample for further 
analysis and are shown as the red triangles in Figure~\ref{fig:mag_dmag}.
Table~\ref{tab:var_info} summarizes the observation details for all 
these objects, and the per-band, per-epoch photometry is given in 
Table~\ref{tab:phot}. 
In addition, for each source we convert the observed bands to rest-frame effective wavelengths 
and the observed time baselines to rest-frame days, 
which are also given in Table~\ref{tab:phot}. 
In addition, we also found two transients that have 
spectroscopic identifications, which are presented in 
Appendix~\ref{appendix:transients}. 

\subsubsection{Notes on Individual Variables} 

   Among the six variables in the spectroscopic sample, five were directly
observed by NIRSpec (the MSA slits were on these variables) and one had
spectroscopy from 10K-DEIMOS on the host galaxy. 
  
\begin{figure*}[hbt!]
    \centering
    \includegraphics[width=0.9\textwidth,height=0.9\textheight,keepaspectratio]{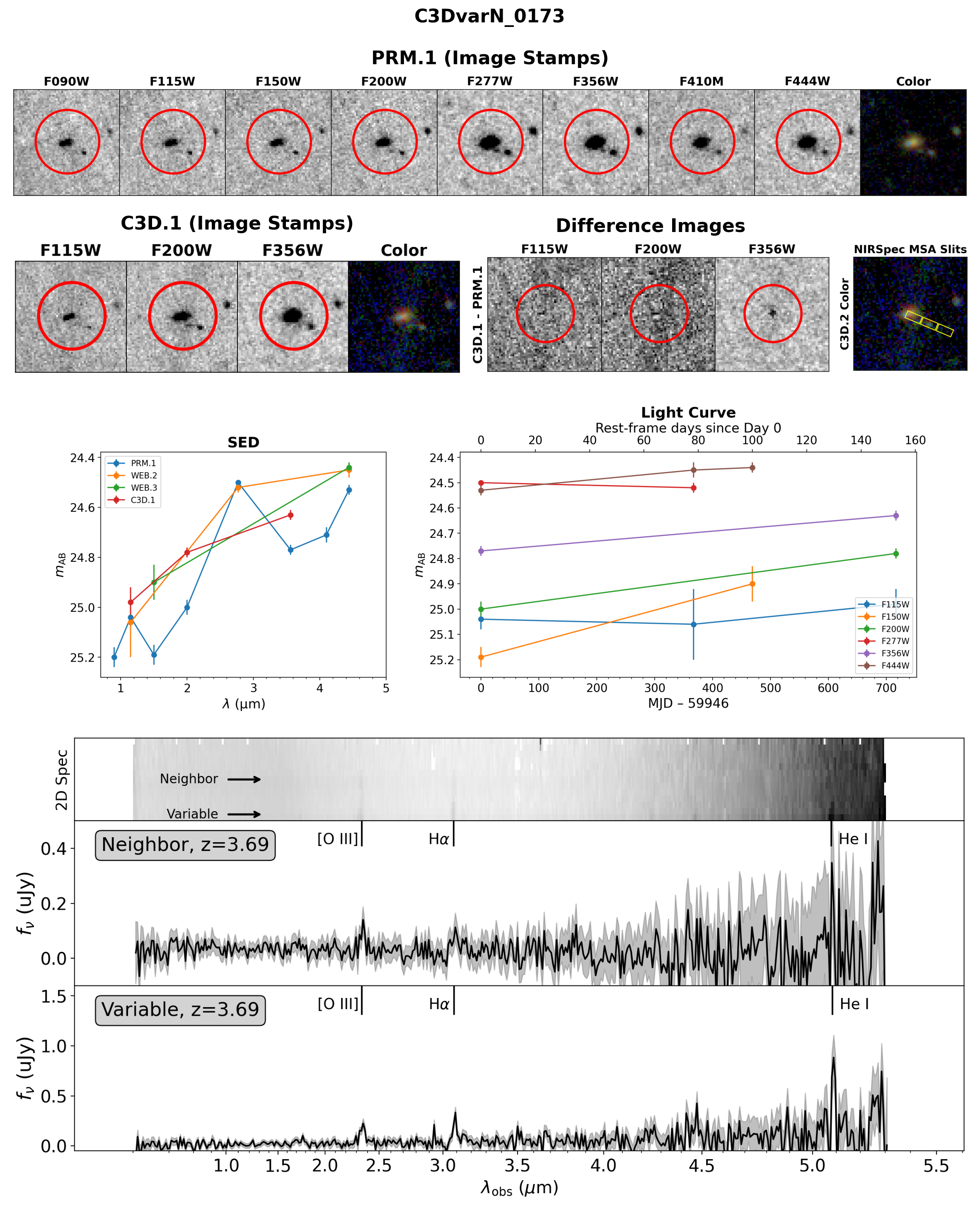}
    \caption{Summary figure for \texttt{C3DvarN\_0173}. The first row shows its
    image stamps (3\farcs6 on a side; North is up and East is to left) in the 
    reference epoch (PRM.1) in all bands and their color composite.
    The second row shows the image stamps (same size and orientation as 
    above) and their color composite in the discovery epoch (C3D.1) in the
    left panels, the difference images between the discovery and reference 
    epochs in F115W, F200W, and F356W (C3D.1$-$PRM.1) in the middle panels,
    and the NIRSpec MSA slit placement superposed on the color composite in 
    the right panel. %The slit covered both the variable and its faint neighbor.
    The third row shows the SEDs of the variable in all epochs (left) and 
    the light curves in the bands that have more than one epoch of photometry (right). 
    The fourth row displays the NIRSpec 2D and 1D spectra. The traces of
    the variable and the neighbor are labeled in the 2D spectrum. The detected
    emission lines are marked on the 1D spectra for both. 
    }
    \label{fig:n0173}
\end{figure*}

\noindent (1) {\tt C3DvarN\_0173}: 
From its images shown in Figure~\ref{fig:n0173}, this object is a disk-like 
galaxy. The NIRSpec slit also covered a faint neighbor $\sim$0\farcs7 away, 
leading to two sets of emission lines on the same individual exposures at 
different vertical (y-axis on the 2D spectrum) positions. For {\sc msaexp} to
successfully extract the two spectra, we manually applied a fixed pixel offset 
on the trace: we summed the 2D spectral data over a narrow y-axis window and 
estimated the local background level using the empty rows within the 2D 
spectrum. Based on their relative positions within the configuration, we were 
able to determine which set of lines belongs to which object. As it turns out, 
both are at the same redshift of $z=3.69\pm0.01$. In other words, the two form a 
close pair separated by only 5.1~kpc.

The variable object has $m_{200}=25.00\pm 0.03$~mag in the reference epoch 
(PRM.1), which corresponds to $M_B\approx -20.9$~mag. Between Day 0 (PRM.1) and 
Day 717 (C3D.1, rest-frame Day 153), it brightened by $0.14$~mag in F356W and $0.22$~mag 
in F200W; however, it brightened by $0.29$~mag in F150W in just 469 days. 
The change in F115W is much smaller ($0.08$~mag between the WEB.2 and C3D.1 epochs). 
%{\bf We used the PSF fitting utility in {\sc Photutils} to determine the variable component, 
%and the measured offset from the host nucleus is $<0.5$~pixel ($<0.2$~kpc).
%}

%Considering this, this object appears to vary more strongly at bluer wavelengths.
%The SED still peaks in F277W, where the [O~\Romannum{3}] emission dominates
%and shows no variability.
%There are two high S/N emission lines ([O~\Romannum{3}] and $\rm H\alpha$) revealed by its spectrum,
%and both turns out to be narrow lines.
%Therefore, two most plausible explanation for the variation
%is (1) a Type II AGN at the center of the galaxy,
%or (2) a deeply embedded transient.

\begin{figure*}[hbt!]
    \centering
    \includegraphics[width=0.9\textwidth,height=0.9\textheight,keepaspectratio]{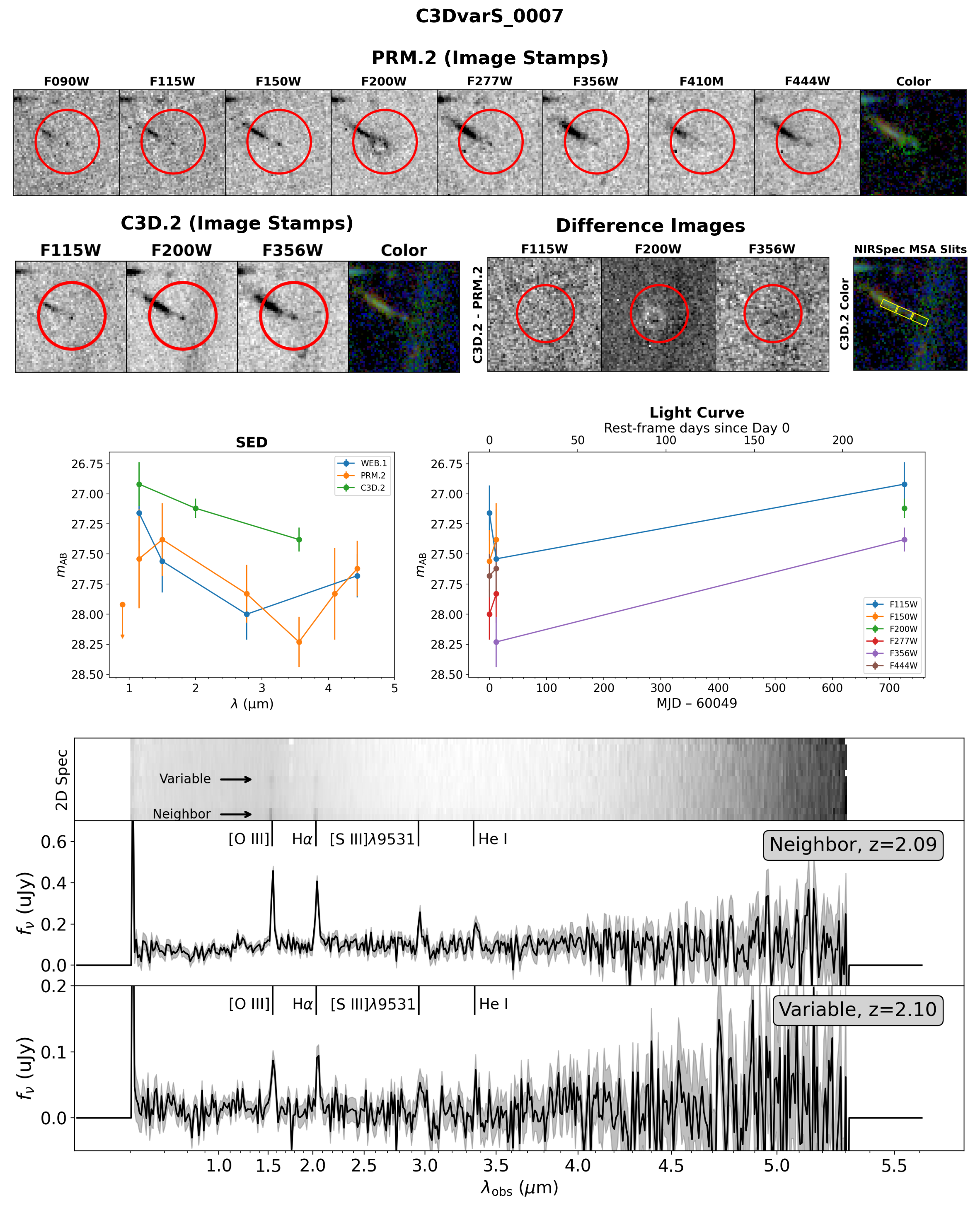}
    \caption{
    Similar to Figure~\ref{fig:n0173} but for \texttt{C3DvarS\_0007}. 
    }
    \label{fig:s0007}
\end{figure*}

\noindent (2) {\tt C3DvarS\_0007}: 
As shown in Figure~\ref{fig:s0007}, this object is faint and very compact, 
which appears as a point-like source in the SW images. In the LW images, its
light is not as concentrated, but this could be due to the lower S/N. 
It lies at the tip of a close neighbor, which is an edge-on disk galaxy whose
centroid is only $\sim$0\farcs9 away. The NIRSpec slit covered both objects, and 
we encountered the same problem of seeing two sets of emission lines in the 
individual exposures as in the case of {\tt C3DVarN\_0173}. We used the same
approach to extract the two objects separately. Based on multiple emission 
lines, the variable and its neighbor are at $z=2.10\pm 0.02$ 
and $z=2.09\pm 0.02$, respectively, and therefore these two objects also form a 
close pair (centroid separation of $\sim$8.4~kpc).
The variable brightened by $0.85$~mag in F356W between PRM.2 (Day 12) and 
C3D.2 (Day 726), with concurrent brightening of $0.62$~mag in F115W. Its PRM.2 
F200W image was severely contaminated by an artifact (known as a 
``snowball'' in the SW detectors), and we had to abandon the photometry in 
this image. Its $m_{150}=27.38\pm0.3$~mag in PRM.2 corresponds to 
$M_B\approx -17.5$~mag when in its quiescent phase.
%{\bf The signal in the difference images are weak S/N$<3$, 
%and thus precise centroid determination is not feasible. 
%Nevertheless, this object is still consistent with being nuclear origin 
%due to its compactness. 
%}
%Because the variable was already clearly present (S/N~$\simeq7$ in F356W) in
%the reference epoch (PRM.2),
%it is less likely a short-lived transient such as a supernova.
%Therefore, a more plausible interpretation is a Type II AGN whose accretion
%was already ongoing before PRM.2 and continued to strengthen through C3D.2.

\begin{figure*}[hbt!]
    \centering
    \includegraphics[width=0.9\textwidth,height=0.9\textheight,keepaspectratio]{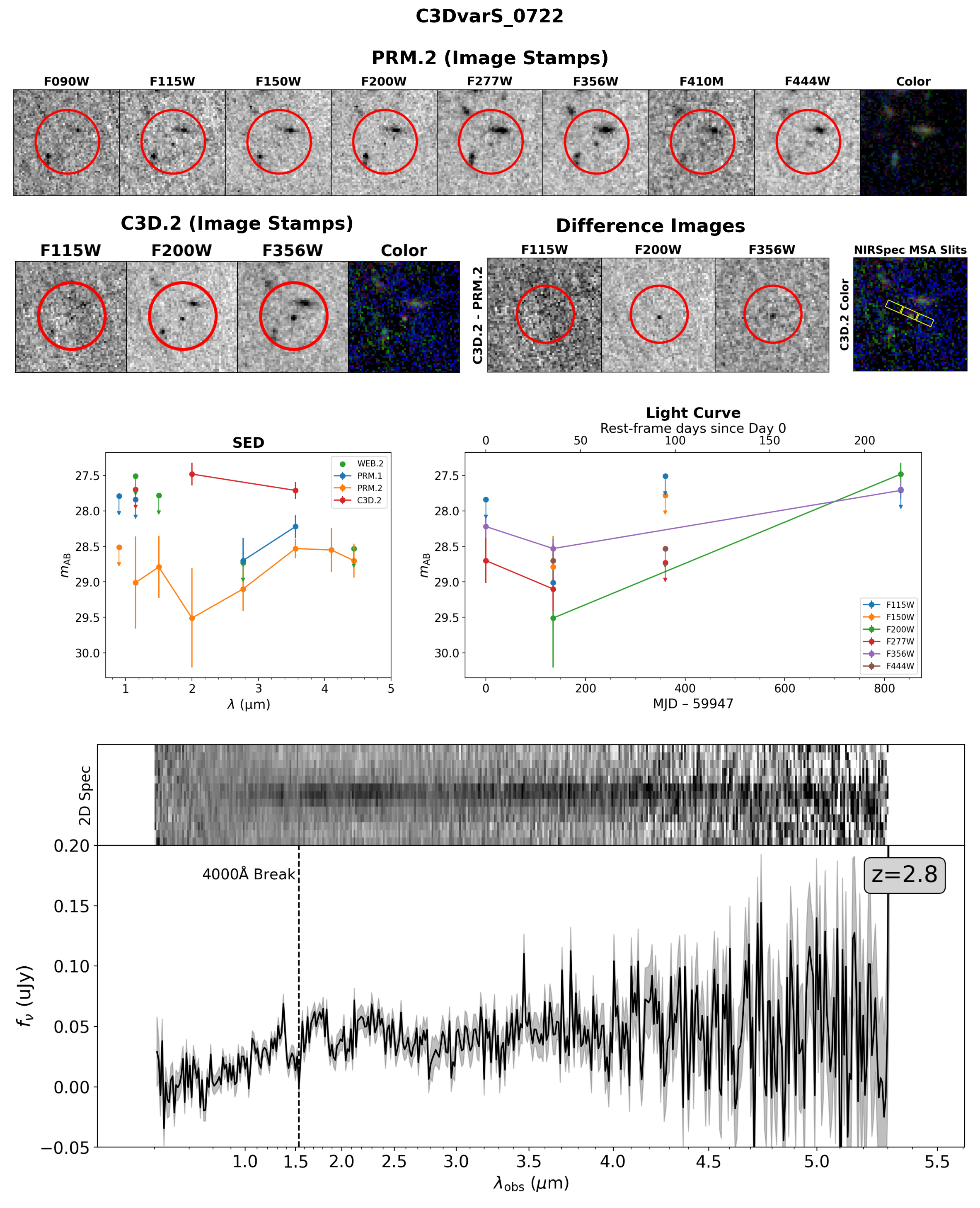}
    \caption{Similar to Figure~\ref{fig:n0173} but for \texttt{C3DvarS\_0722}.
    This object does not have any emission lines detected in the spectrum. 
    However, there is a strong absorption feature at $\sim$1.5~$\mu$m, which we
    tentatively identify as the 4000~\AA\ break. With this identification, the
    source is at $z=2.8$.
    }
    \label{fig:s0722}
\end{figure*}

\begin{figure*}[hbt!]
    \centering
    \includegraphics[width=0.9\textwidth,height=0.9\textheight,keepaspectratio]{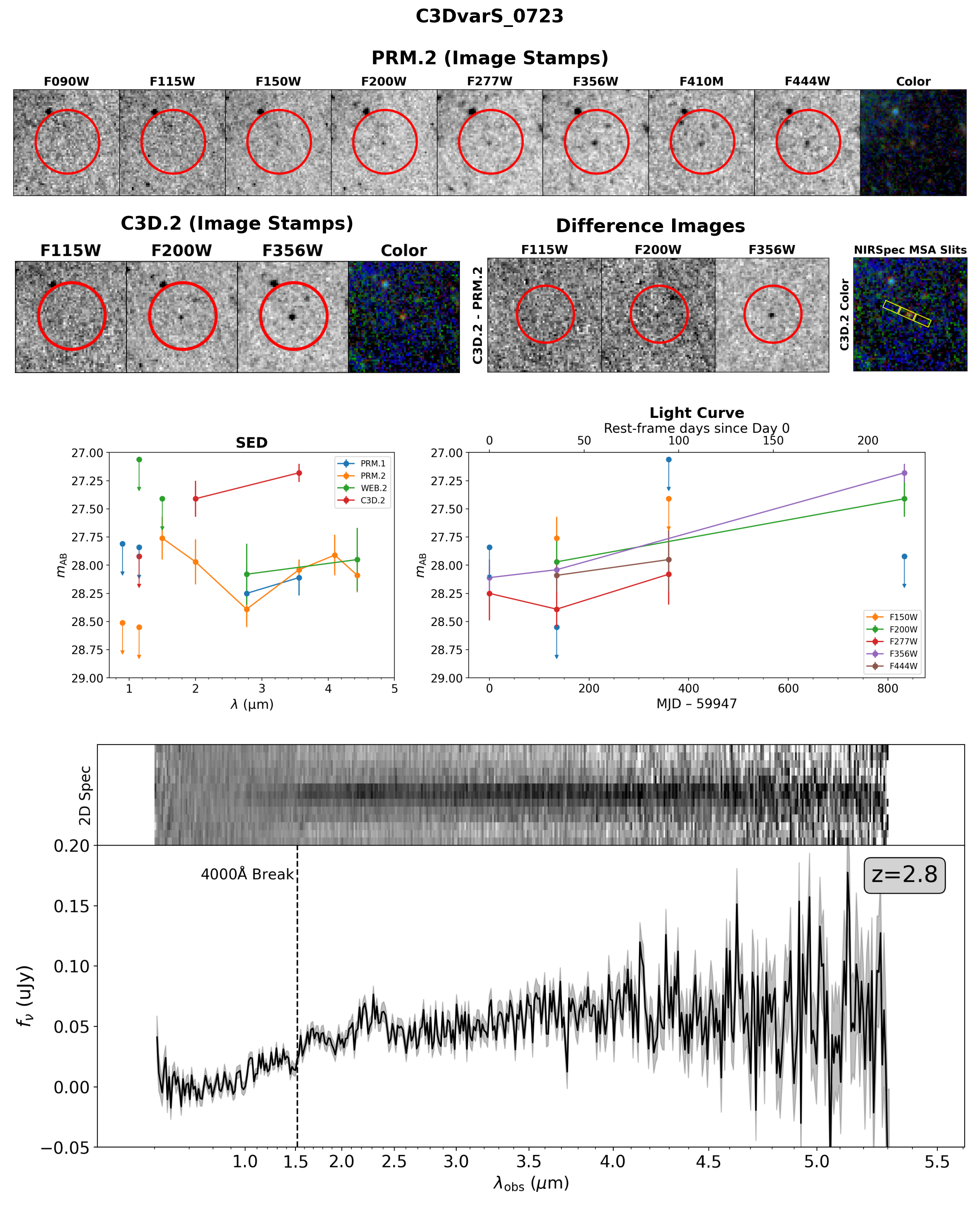}
    \caption{Similar to Figure~\ref{fig:s0722} but for \texttt{C3DvarS\_0723}.
    Its redshift is also based on the absorption feature identified as the
    4000\AA\ break.
    }
    \label{fig:s0723}
\end{figure*}

\noindent (3),(4) {\tt C3DvarS\_0722 \& C3DvarS\_0723}: 
These two variables, presented in Figures~\ref{fig:s0722} and 
\ref{fig:s0723}, form an interesting pair. They are discussed together because
they share a lot of similarities. They are only 2\farcs70 apart, and
both are point-like. Over 698 days between PRM.2 and C3D.2, 
they brightened by $0.82$~mag and $0.86$~mag in F356W, respectively. 
In the PRM.2 epoch, both showed a ``V-shaped'' SED in $f_\nu$ with a trough at 
$2-3\mu$m in the observed frame, which is similar to those of the so-called 
``Little Red Dots'' (LRDs). However, they do not meet the color criteria 
employed by most of the recent LRD selections
\citep[e.g.,][]{Perez-Gonzalez2024,Barro2024,Labbe2025} mainly because their 
SEDs have a shallower rise from 2 to 5~$\mu$m; for example, they have 
$m_{277}-m_{444}=0.4$ and 0.3~mag, respectively, as oppose to the usual 
$m_{277}-m_{444}\geq 1.0$~mag criterion.
Their spectra (taken on the same day as the C3D.2 imaging) reveal a break at 
$\sim$1.5~$\mu$m but no high S/N emission lines. We tentatively identify this 
break as the 4000\AA\ break, which gives $z\approx2.8$ for both objects.

\begin{figure*}[hbt!]
    \centering
    \includegraphics[width=0.9\textwidth,height=0.9\textheight,keepaspectratio]{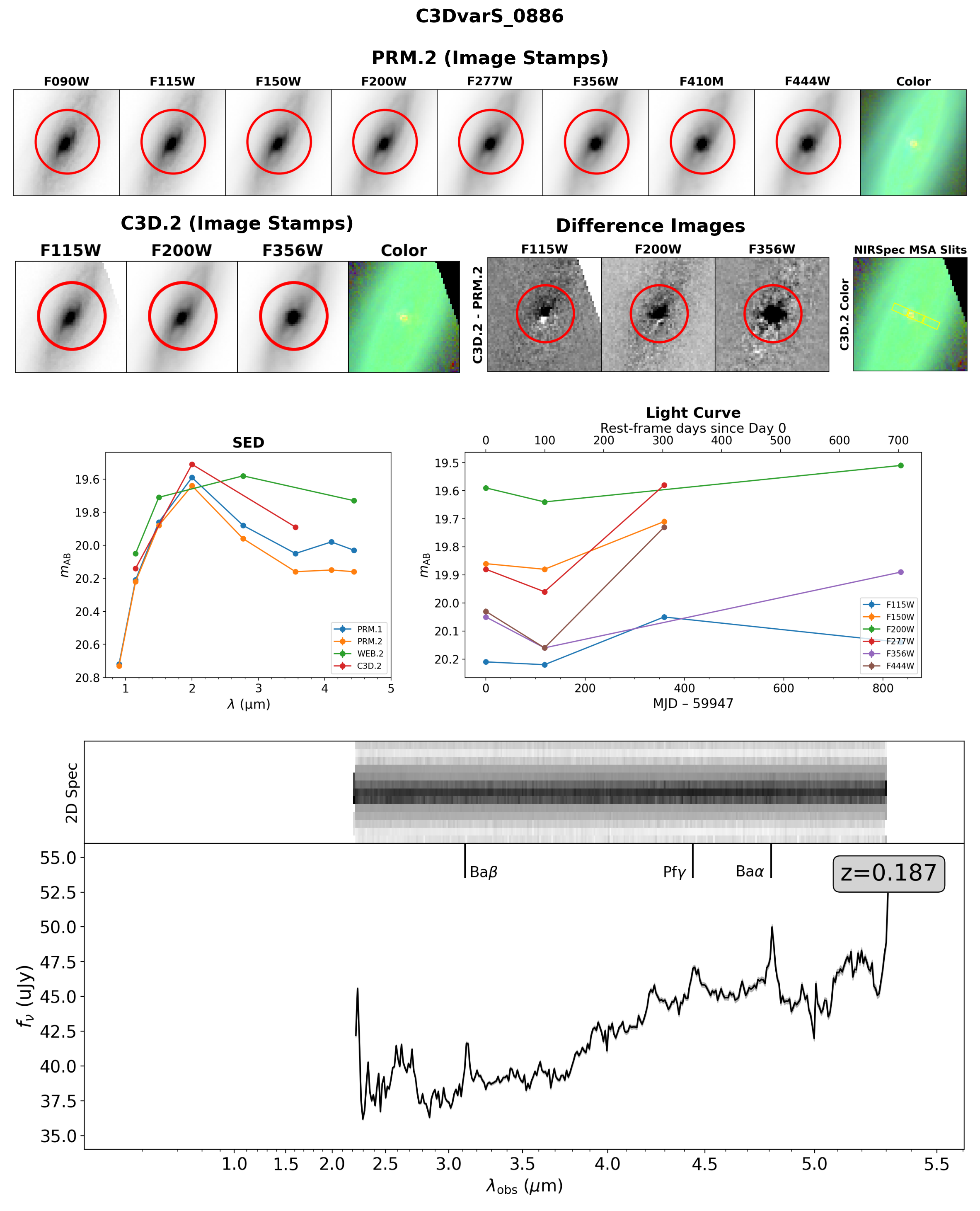}
    \caption{Similar to Figure~\ref{fig:n0173} but for \texttt{C3DvarN\_0886}. 
    The lack of data at $\lambda_{\rm obs}\lesssim2.2\mu$m is because the
    spectrum falls in the detector gap.}
    \label{fig:s0886}
\end{figure*}

\noindent (5) {\tt C3DvarS\_0886}: 
This object is a large spiral galaxy, and the variability happened in its 
nucleus, which can be seen from the difference images
(Figure~\ref{fig:s0886}). It has a previously reported redshift of $z=0.1854$
from the 3D-HST survey \citep[][]{Brammer2012_3dhst}.
The NIRSpec slit covered exactly the nuclear region,
and the spectrum reveals three prominent lines, Ba$\beta$, Pa$\gamma$ and
Ba$\alpha$, which give $z=0.187\pm 0.004$.
It was observed in all four epochs by NIRCam, and its light curve is special 
among our sample: it traces a bright-dim-brighter sequence over the four epochs 
spanning a total of 836 days. The MAG\_ISO magnitudes only vary at a few
percent level (but still passing our selection criterion as described in
Section~\ref{sec:var_search}), which is understandable because these measure 
the brightness of the whole galaxy. To better reflect the actual change in its 
nucleus, we adopted the MAG\_AP magnitudes measured in $r=0\farcs24$ for this
object, which are reported in Table~\ref{tab:var_info}.
In F356W, we observed a peak-to-valley change of $\Delta m_{356}=0.27$~mag. 
Its variation between the brightening epochs appears to be more significant in
the redder passbands: 
between PRM.2 (Day 118) and WEB.2 (Day 359), 
the variable brightened by $0.17$~mag in both F115W and F150W, 
and by $0.38$ and $0.43$~mag in F277W and F444W, respectively. 

%Again, no broad wings were observed. 
%At this low redshift, the entire observing window corresponds to $\sim670$ days in the rest frame, 
%implying that each variability period lasts $\sim500$ days. 

\begin{figure*}[hbt!]
    \centering
    \includegraphics[width=0.9\textwidth,height=0.9\textheight,keepaspectratio]{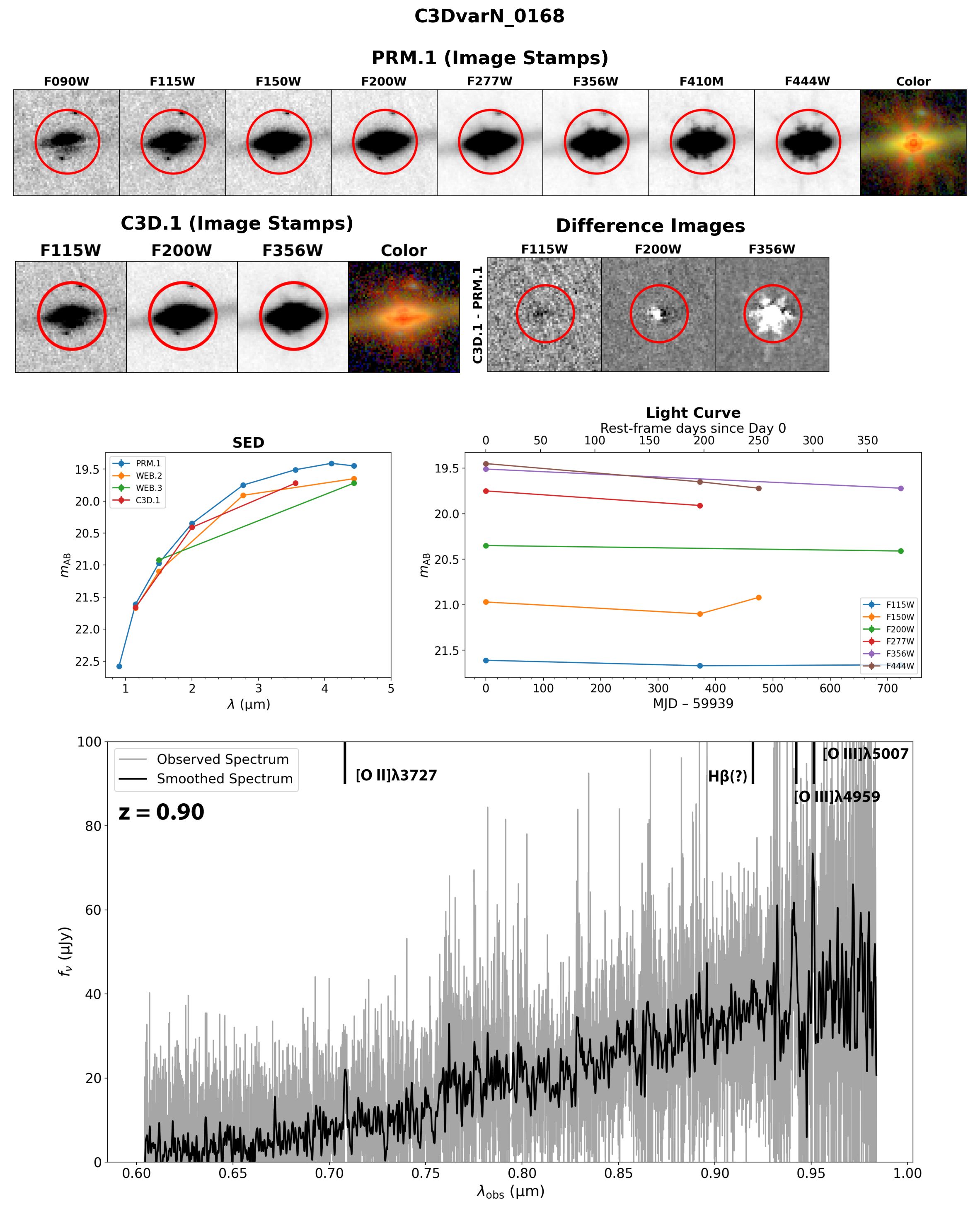}
    \caption{
    Similar to Figure~\ref{fig:n0173} but for \texttt{C3DvarN\_0168}. It was not
    observed by NIRSpec, and the spectrum show in the bottom is from the Keck 
    10K-DEIMOS survey, which ranks its redshift as secure (Qf=4). 
    The original and smoothed spectra are shown in gray and black, respectively, 
    and the [O~\Romannum{2}] line and the [O~\Romannum{3}] doublet are labeled.
    The H$\beta$ line is not visible at the expected wavelength.
    }
    \label{fig:n0168}
\end{figure*}

\noindent (6) {\tt C3DvarN\_0168}: 
Figure~\ref{fig:n0168} presents this object, which does not have NIRSpec
observations but has $z=0.90$ as reported by the 10K-DEIMOS
survey. It is an edge-on disk galaxy that has a dust lane, which is the most 
visible in the SW bands (especially in F090W). It has a central point source 
dominant in the LW bands, which must be an AGN heavily obscured by the dust of
the host galaxy. The variations indeed happened 
in the central region, which is obvious from the difference images.
The source faded by $\Delta m_{356}=0.21$~mag (as measured by MAG\_ISO)
between the PRM.1 (Day 0) and C3D.1 (Day 723), while the changes in the bluer 
F115W and F200W bands were not significant. Interestingly, it brightened in 
F150W (rest-frame $\sim7900{\rm \AA}$) by 0.18~mag between the WEB.2 (Day 373)
and WEB.3 (Day 475) epochs. 
The original 1D spectrum retrieved from 10K-DEIMOS was very noisy, and we 
smoothed it with a 1D Gaussian kernel in order to show the detected lines. The 
source position reported by 10K-DEIMOS is slightly different from the NIRCam
centroid, which only amounts to 0\farcs2 and is likely due to the systematic 
offset between the world coordinate systems adopted. We believe that the 
10K-DEIMOS slit was placed at the center of this galaxy.

\begin{table*}[hbt!]
    \raggedright
    \caption{Emission line properties for five variables measured by NIRSpec}
    \resizebox{\textwidth}{!}{
    \begin{tabular}{lcccccccccc} \hline 
        SID & [O~\Romannum{2}]$\lambda$3727 & [O~\Romannum{3}]$\lambda\lambda$4959,5007 & $\rm H\alpha$ & [S~\Romannum{2}]$\lambda6716$ & [S~\Romannum{3}]$\lambda9531$ & He I $\lambda10830$ & $\rm Pa\beta$ & $\rm Ba\beta$ & $\rm Pf\gamma$ & $\rm Ba\alpha$ \\ \hline 
        {\bf C3DvarN\_0173} & $<6.73$ & $4.88\pm1.66$ & $3.27\pm0.74$ & $0.87\pm0.63$ & $<2.83$ & $1.44\pm0.57$ & ... & ... & ... & ... \\ 
         $\Delta\lambda_{\rm obs}$ ($\Delta\lambda_{\rm inst}$) & - & $526\pm138$ ($286\pm4$) & $238\pm45$ ($216\pm2$) & $323\pm190$ ($211\pm2$) & - & $136\pm40$ ($146\pm8$) & ... & ... & ... & ... \\ 
         %$\Delta v_{\rm rest}$ & - & $1432.48\pm375.27$ & $675.68\pm110.65$ & - & - & $171.43\pm50.67$ & ... & ... & ... & ... \\ \hline 
         $\Delta v$ & - & - & $973\pm1044$ & $2328\pm2389$ & - & $<355$ & ... & ... & ... & ... \\ \hline 
        {\bf C3DvarS\_0007} & $<5.72$ & $4.72\pm1.40$ & $2.49\pm0.82$ & - & $<1.36$ & $<1.08$ & $<1.07$ & ... & ... & ... \\ 
         $\Delta\lambda_{\rm obs}$ ($\Delta\lambda_{\rm inst}$) & - & $464\pm22$ ($345\pm15$) & $322\pm17$ ($326\pm6$) & - & - & - & - & ... & ... & ... \\ 
         %& - & $2896.5\pm135.5$ & $1528.8\pm80.3$ & - & - & - & - & ... & ... & ... \\ 
         $\Delta v$ & - & - & $<1388$ & - & - & - & - & ... & ... & ... \\ \hline
        %{\bf Neighbor} & $<8.95$ & $15.87\pm1.77$ & $8.89\pm1.23$ & - & $1.33\pm0.46$ & $1.07\pm0.46$ & $<1.65$ & ... & ... & ... \\ 
         %$\Delta\lambda_{\rm obs}$ ($\Delta\lambda_{\rm inst}$) & - & $316\pm27$ ($346\pm15$) & $372\pm40$ ($325\pm6$) & - & $225\pm59$ ($225\pm2$) & $338\pm112$ ($198\pm3$) & - & ... & ... & ... \\ 
         %& - & $1968.0\pm167.9$ & $1766.2\pm190.3$ & - & $736.7\pm193.8$ & $973.2\pm321.6$ & - & ... & ... & ... \\ \hline 
         % $\Delta v$ & - & ... & $2667\pm1222$ & - & $<524$ & $2446\pm1234$ & - & ... & ... & ... \\ \hline 
         %C3DvarS\_0187 & $7.04\pm1.94$ & $15.07\pm1.44$ & $13.29\pm0.94$ & $0.92\pm0.31$ & $2.27\pm0.78$ & $<1.99$ & $0.84\pm0.23$ & ... & ... & ... \\ 
         % & $2289.2\pm465.7$ & $2230.7\pm157.8$ & $1681.9\pm249.4$ & $601.5\pm317.4$ & $747.2\pm199.4$ & - & $387.3\pm83.5$ \\ \hline 
        {\bf C3DvarS\_0722} & $<2.55$ & $<1.55$ & $<0.97$ & $<0.92$ & $<0.61$ & $<0.62$ & $<0.84$ & ... & ... & ... \\
         & - & - & - & - & - & - & - & ... & ... & ... \\ \hline 
        {\bf C3DvarS\_0723} & $<1.71$ & $<1.09$ & $<0.71$ & $<0.65$ & $<0.41$ & $<0.44$ & $<0.59$ & ... & ... & ... \\ 
         & - & - & - & - & - & - & - & ... & ... & ... \\ \hline 
        {\bf C3DvarS\_0886} & ... & ... & ... & ... & ... & ... & ... & $40.78\pm1.62$ & $19.15\pm1.78$ & $17.90\pm0.97$ \\ 
         $\Delta\lambda_{\rm obs}$ ($\Delta\lambda_{\rm inst}$) & ... & ... & ... & ... & ... & ... & ... & $352\pm11$ ($213\pm2$) & $498\pm41$ ($160\pm3$) & $331\pm14$ ($152\pm5$) \\ 
         %& ... & ... & ... & ... & ... & ... & ... & $2855.8\pm86.1$ & $2829.0\pm233.2$ & $1739.7\pm73.8$ \\ \hline 
         $\Delta v$ & ... & ... & ... & ... & ... & ... & ... & $2696\pm134$ & $3184\pm292$ & $1833\pm100$ \\ \hline 
    \end{tabular}
    }
    \tablecomments{
    The first row of each target reports the line intensities in the 
    units of $10^{-18}$~erg~cm$^{-2}$~s$^{-1}$. For non-detections, the 2$\sigma$
    upper limits are quoted. If a line is not covered within the wavelength 
    range of the spectrum, it is labeled as ``...".
    The second row gives the line widths (FWHM) in the unit of \AA, together with
    the instrumental line broadening factors at the line locations (in 
    parenthesis) in the same unit.
    The third row (not present for \texttt{C3DvarS\_0722} and 
    \texttt{C3DvarS\_0723} that do not have emission lines) lists the 
    corresponding rest-frame intrinsic line widths after correcting the 
    instrumental line broadening, which are quoted in velocities in the units of 
    km~s$^{-1}$. Upper limits are given when the measured FWHM values are
    narrower than the line broadening factor (see text for details). Velocities
    are not calculated for the [O~\Romannum{3}] doublet as these would be 
    meaningless due to the line blending.
    }
    \label{tab:line_measure}
\end{table*}

\begin{figure*}
    \centering
    \includegraphics[width=\textwidth,height=0.9\textheight,keepaspectratio]{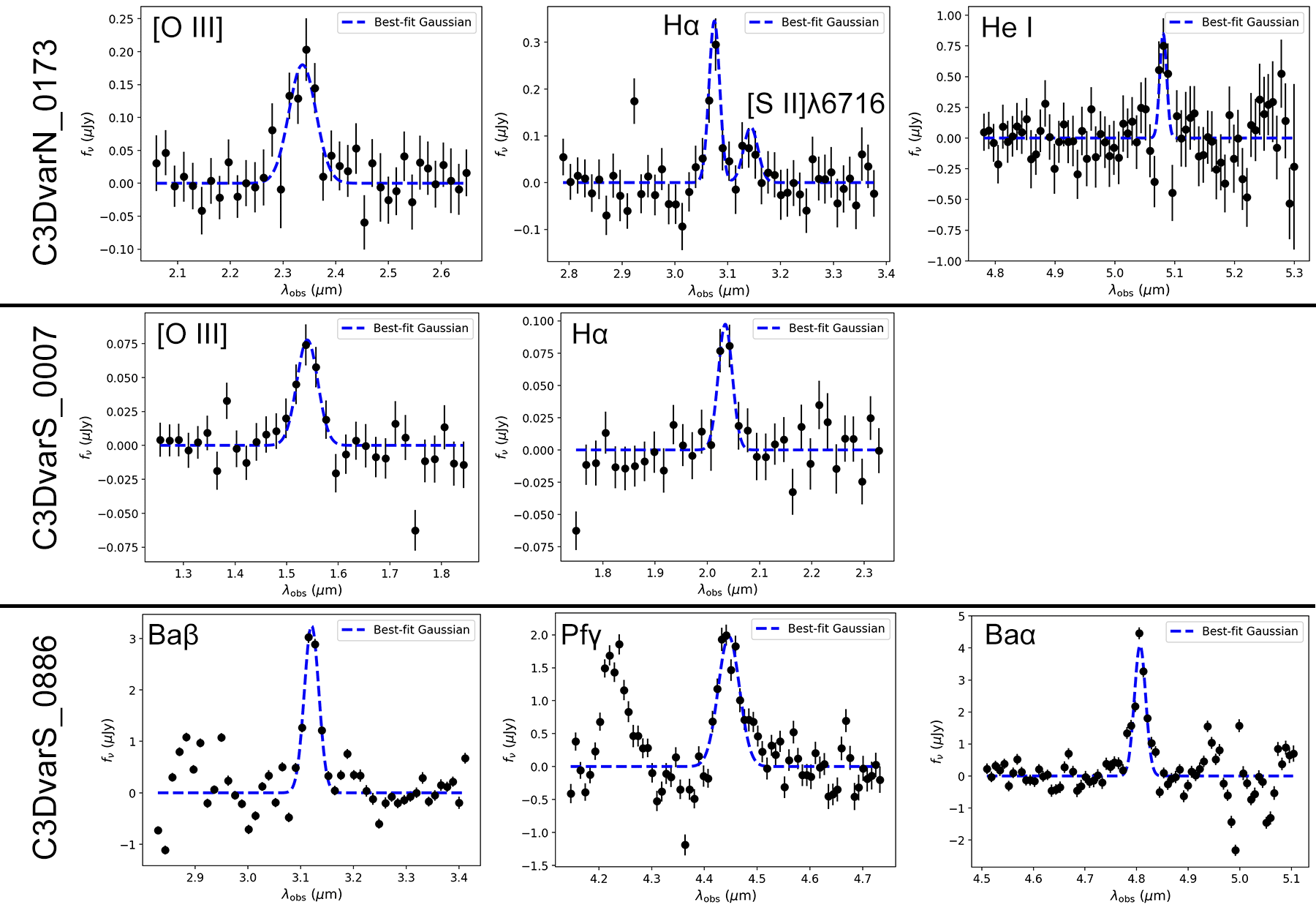}
    \caption{Fitting of the emission lines reported in 
    Table~\ref{tab:line_measure}. In each panel, the continuum-subtracted 
    spectrum around each emission line is shown as the black circles with error bars, and the best-fit Gaussian profile is shown in blue. 
    }
    \label{fig:line_fit}
\end{figure*}

   Appendix~\ref{appendix:altplot} shows the SEDs of these six variables in 
terms of monochromatic luminosity ($\nu L_\nu$) as well as their ``light curves''
as measured in the difference images.

\subsubsection{Emission line measurements}

    For the three objects that have emission lines detected in the NIRSpec data,
we measured their line properties following the methodology in 
\citet[][]{SY2025b}. Briefly, we fitted the continuum of a spectrum with a 
seventh-order Chebyshev polynomial using the {\tt fit\_generic\_continuum} 
utility in the {\sc astropy/specutils} package \citep[][]{Earl24_specutils}. 
When performing the fit, we excluded the wavelength regions where the emission 
lines fall onto. The fitted continuum was then subtracted from the spectrum for
line measurement. {\tt C3DvarS\_0886} required some special treatments. Its
continuum varies greatly with wavelength and could not be fitted well by a 
polynomial. We ended it up with fitting a three-degree spline using the
{\tt LSQUnivariateSpline} function in the {\sc scipy} package \citep[][]{Virtanen2020_scipy}. 

    After the continuum subtraction, we fitted a Gaussian profile to each 
emission line. We adopt the full-width-at-half-maximum (FWHM) values as the line
widths, and the line fluxes were measured within $2\times{\rm FWHM}$ of the 
fitted central wavelengths. The emission line fits are presented in 
Figure~\ref{fig:line_fit}, and the measurement results are summarized in 
Table~\ref{tab:line_measure}. 

    It is important to note that the observed line widths (FWHM values reported
as ``${\rm \Delta\lambda_{obs}}$'' in Table~\ref{tab:line_measure}) are all 
severely impacted by the instrumental line broadening due to the very low 
resolution of the PRISM mode. Strictly speaking, such a low-resolution mode is 
not suitable for line width measurements. Nevertheless, the widths obtained 
using these available data, albeit with large uncertainties, still provide 
critical information for our analysis. 

    The currently available NIRSpec resolving power curves at the
JWST User Documentation (JDOC) site
\footnote{\href{https://jwst-docs.stsci.edu/jwst-near-infrared-spectrograph/nirspec-instrumentation/nirspec-dispersers-and-filters}{https://jwst-docs.stsci.edu/jwst-near-infrared-spectrograph/nirspec-instrumentation/nirspec-dispersers-and-filters}.} 
are based on the pre-launch estimates. In fact, the in-flight NIRSpec 
performances are better than these curves indicate. For the purpose of this 
study, we estimated the actual resolving power of the PRISM mode, which is 
detailed in Appendix~\ref{appendix:pr}.
Using the new resolving power curve, we derived the instrumental broadening 
factor (${\rm FWHM_{pb}}$) for each measured line, which is reported in 
parenthesis in Table~\ref{tab:line_measure} as ``${\rm\Delta\lambda_{inst}}$''. 
As explained in Appendix~\ref{appendix:pr}, the new resolving power curve could 
slightly overestimate the actual performance; this can be seen in the 
[He~\Romannum{1}] line of \texttt{C3DvarN\_0173} and the H$\alpha$ line of
\texttt{C3DVarS\_0007}, where the measured widths are 
smaller than the broadening factors (but still within the uncertainties).

   From the observed line widths (${\rm FWHM_{obs}}$), we calculate the true 
line widths (${\rm FWHM_t}$) by subtracting the broadening factor 
(${\rm FWHM_{pb}}$) in quadrature, i.e.,

\begin{align}
    {\rm FWHM_t=\sqrt{\rm FWHM_{obs}^2-FWHM_{pb}^2}},
\end{align}

\noindent which is then converted to velocities by 
$\Delta v =c\times ({\rm FWHM_t}/\lambda_c)$, where $\lambda_c$ is the line 
center and $c$ is the speed of light. These velocities are also reported in 
Table~\ref{tab:line_measure}. Note that the [O~\Romannum{3}] doublet 
measurements are not converted to velocities because the wavelength difference 
between the two lines (48\AA\ in rest frame) cannot be separated due the line 
broadening.

\section{Discussion}

The conventional wisdom would readily attribute the variability of galaxies to 
the AGNs that they host. However, our spectroscopic sample suggests a more subtle 
picture. On the one hand, the variability all happened in the nuclear region 
of these objects, which is in line with the AGN interpretation. Using the PSF 
fitting utility in {\sc Photutils}, we determined the centroid of the variable 
component in the F356W difference image for each source and found that the measured 
offset from the host nucleus is $<30$~mas for all objects except 
{\tt C3DvarS\_0007}, which is too faint in the difference image to obtain reliable 
centroids. Nevertheless, since {\tt C3DvarS\_0007} is very compact, there is no 
evidence arguing against that its variability originated from its nucleus. 
On the other hand, most of the objects in our spectroscopic sample show some
characteristics that are not fully consistent with the conventional AGN variability
picture. These are discussed in detail below. 

\subsection{Type~1 AGNs or not}

   First of all, {\tt C3DvarN\_0168} ($z=0.90$) and {\tt C3DvarS\_0886}
($z=0.187$) do seem to host an AGN: (1) the strong central point source in the 
former can hardly be explained by anything else, and (2) their brightness 
variations indeed happened in their nuclei, as the difference images show. In 
addition, both objects are detected in X-ray by the Chandra Cosmos Legacy Survey 
\citep[][]{Civano2016_chandra-cosmos}, and {\tt C3DvarN\_0168} is also detected 
in radio by the VLA-COSMOS 3~GHz survey \citep[][]{Smolcic_vla3g}. 
Their X-ray fluxes over 0.5--2~keV are $(3.74\pm0.34)\times10^{-15}$ 
and $(1.81\pm0.07)\times10^{-14}$~erg~s$^{-1}$~cm$^{-2}$ 
for {\tt C3DvarN\_0168} and {\tt C3DvarS\_0886}, respectively; 
the 3~GHz flux density of the latter is $11.6\pm2.4$~$\mu$Jy. 
 
   As mentioned in Section~\ref{sec:intro}, the unification model implies that
type~2 AGNs should rarely vary, and therefore one would expect that our variable 
objects are type~1 AGNs. Due to the poor spectral resolution, it is not possible 
to separate the broad and narrow components of the permitted lines in our data. 
Nevertheless, the three permitted lines of \texttt{C3DvarS\_0886} all have 
$\Delta v$ exceeding 1000~km~s$^{-1}$, which is consistent with being a type~1 
AGN. The differences in the inferred velocities could be attributed to the 
errors in the resolving power estimates. The interpretation of 
\texttt{C3DvarN\_0168}, on the other hand, is not straightforward. Its optical 
spectrum from the 10K-DEIMOS survey does not detect the H$\beta$ emission line 
(despite that it is within the wavelength coverage), and therefore we cannot 
determine its type based on the available spectroscopy (in fact, it is not 
classified as an AGN in 10K-DEIMOS). However, it is an edge-on disk galaxy with 
an inclination nearly perpendicular to the sight line. While it is still a 
matter of debate whether the torus of an AGN is co-planar with its host galaxy 
\citep[see e.g.,][]{Hopkins2012}, it has been shown that very few type~1 AGNs 
are found in edge-on disk galaxies
\citep[e.g.,][]{Keel1980, Maiolino1995, Lagos2011, Gkini2021}. 

   {\tt C3DvarN\_0173} ($z=3.69$) and {\tt C3DvarS\_0007} ($z=2.09$) are more
complicated. It is interesting to note that both are in a close-pair environment 
and that the other member is not a variable. In both cases, neither the source 
nor the neighbor is detected in the Chandra Cosmos Legacy Survey (limiting depth 
of $2.7\times 10^{-16}$~erg~s$^{-1}$~cm$^{-2}$ over 0.5--2~keV) or in the 
VLA-COSMOS 3~GHz survey (RMS noise of 2.3~$\mu$Jy~beam$^{-1}$).
For {\tt C3DvarN\_0173}, its He~\Romannum{1}$\lambda$10830 line has the measured 
width ($136\pm40$~\AA) smaller than the broadening factor ($146\pm 8$~\AA), 
which is most likely due to an overestimated resolving power (see Section~3.2.2 
and Appendix~\ref{appendix:pr}). If we assume that the resolving power is off by 
3~$\sigma$ and take the 3~$\sigma$ broadening factor lower limit (122~\AA), the 
inferred velocity upper limit would be $\Delta v < 355$~km~s$^{-1}$. The 
velocity based on its H$\alpha$ line is$\Delta v = 973$~km~s$^{-1}$ but with an 
uncertainty of 1044~km~s$^{-1}$. Such a large uncertainty is mostly due to the 
large error in the width measurement ($\pm 45$~\AA), which is unavoidable 
because of the poor spectral resolution: at this wavelength ($\sim$3.08~$\mu$m), 
one pixel corresponds to $\sim$110~\AA. More importantly, the H$\alpha$ line 
would be blended with [N~\Romannum{2}]$\lambda\lambda$6548,6583 if these weaker 
lines are present; the blending with [N~\Romannum{2}]$\lambda$6583 alone would 
easily make a combined ``single line'' of a width exceeding $\sim$235~\AA. 
Considering all this, we conclude that the current PRISM data do not show 
strong evidence that C3DvarN\_0173 hosts a type~1 AGN. 
%Considering all
%these, we conclude that there is no strong evidence supporting
%{\tt C3DvarN\_0173} being a type~1 AGN. 
The situation for {\tt C3DvarS\_0007} is similar. The only permitted line 
detected is H$\alpha$, however, its measured width ($322\pm17$~\AA) is 
smaller than the broadening factor ($326\pm 6$~\AA). Applying the same 
reasoning as above, we obtain the 3~$\sigma$ velocity upper limit of 
$\Delta v < 1388$~km~s$^{-1}$. 
In other words, it is also consistent with a narrow-line object, and there 
is no strong evidence supporting its being a type~1 AGN. 
On the other hand, we should point out that the current PRISM data of both
objects do not have sufficient S/N to rule out the possibility that they have
a weak broad-line component. This is discussed in detail in 
Appendix~\ref{appendix:linesim}.
Spectroscopy of higher S/N and higher resolution will be needed 
to confirm their narrow-line nature more robustly. 
%Considering the possible line blending,
%there is hardly any supporting evidence for a type~1 AGN. 

   Lastly, {\tt C3DvarS\_0722} and {\tt C3DvarS\_0723} (which form a pair by
themselves) do not show any emission lines. Qualitatively, their lack of 
emission lines is similar to BL Lac objects, which are known to have featureless 
UV/optical spectra and strong variability. 
In the unification model, BL Lac objects should be classified as type 1 because
our view to their accretion disks is not obscured by their dust tori (having
their beamed relativistic jets pointing along the sight lines).
However, these two sources are not likely classic BL Lac objects for two reasons:
(1) the spectra of BL Lac objects can be described by power-law, but those of 
these two objects cannot; (2) classic BL Lac objects are usually loud in radio
and X-ray, but these two sources are not detected in radio or X-ray by the two
aforementioned surveys.
On the other hand, there are indeed some BL-Lac-like objects (i.e., with
featureless UV/optical spectra) to be weak or quiet in radio and X-ray 
\citep[e.g.,][]{Londish2004, Collinge2005, Plotkin2010}, and they might be the
same kind of objects as the so-called ``weak line quasars'' 
\citep[WLQs; se.g.,][]{Fan1999, Fan2006, Shemmer2009, Diamond-Stanic2009,
Meusinger_Balafkan2014}, a rare type of quasars with very weak or no emission 
lines and being weak or quiet in X-ray and radio. 
It is unclear whether a complete obscuration of their line-emitting regions 
by dust could be an explanation, as their continua generally do not show 
obvious sign of reddening \citep[e.g.,][]{Meusinger_Balafkan2014}. 
Recently, \citet[][]{Kumar2025} studied the temporal behavior of a large sample of WLQs 
and confirmed that they clearly show long-term (months to years) variability, 
albeit being weaker than typical type 1 quasars. In this sense, 
{\tt C3DvarS\_0722} and {\tt C3DvarS\_0723} could be similar to WLQs. 
The problem, however, is that they are much less luminous than WLQs: 
at $z=2.8$, their $m_{200}$ magnitudes in the brightened phase (epoch C3D.2) 
correspond to only $M_V\approx -18.0$~mag (as oppose to $M_V\lesssim -23$~mag for
quasars). 

\subsection{Variable Type 2 AGNs or nuclear transients}

    The above analysis shows that only one object (\texttt{C3DvarS\_0886}) in 
our spectroscopic sample (six objects in total) appears to be a type~1 AGN. If
we still attribute the variability in the other five to AGN, they would be
type~2 AGNs. As mentioned in Section~\ref{sec:intro}, low-redshift surveys
indeed show that a small fraction of type~2 AGNs could be variables. However,
such a high fraction of variable type~2 AGNs in our sample would be unusual.
Therefore, we should further examine this interpretation.

   Among these five, the most likely type~2 AGN is \texttt{C3DvarN\_0168}. 
However, this assessment is only based on morphological arguments (a 
prominent nuclear point source in an edge-on disk galaxy), and a definite 
conclusion will have to wait for better spectroscopy in the future, e.g., deep
NIRSpec grating spectroscopy of at least $R\sim 1000$ at $\sim$1.25~$\mu$m to
detect whether its H$\alpha$ line has a broad component. 
%{\bf In addition, its variability is indeed generally more significant in the redder bands, 
%which is consistent with a type 2 picture that the nuclear emission is partially obscured 
%and the variability is thus more easily detected at longer wavelengths; 
%however, such variability behavior may not be unique to type 2 AGNs. }
Determining its type 
will have important implications: if it is type~2, we will add a strong case 
challenging the unification model; if it is type~1, this object will then be an
extreme case where an AGN axis is severely misaligned with the rotation axis of 
its host galaxy.

   \texttt{C3DvarN\_0173} and \texttt{C3DvarN\_0007} also need deeper and higher
resolution spectroscopy to determine whether they are AGNs in the first place. 
While they have multiple lines detected in the current PRISM spectra, these are 
not sufficient for a usual line diagnostics that separates AGNs from 
star-forming emission-line galaxies. For example, \texttt{C3DvarN\_0173} has a
marginal detection of [S~\Romannum{2}]$\lambda$6716, but it lacks the detection
of H$\beta$ because of the low resolution and sensitivity. The low resolution 
even prevents a reliable measurement of the line flux upper limit.

    The most puzzling cases are \texttt{C3DvarS\_0722} and 
\texttt{C3DvarS\_0723}, which do not show any emission lines. This is unlikely to
be explained by some obscuring mechanisms that hide both their BLR and NLR
from view, because the same mechanisms would block the AGN continuum as well.
As discussed in the previous section, they share some similarities with WLQs but 
fall several magnitudes short in luminosity. Here we consider another possibility.
A plausible scenario for variable type~2 AGNs is that they could be CLAGNs whose 
continuum variabilities accompany the disappearance or emergence of the broad 
line components at the time of type transition. The NIRSpec observations of 
these two objects were on the same day as their imaging discovery epoch (C3D.2); 
if they are indeed CLAGNs, this would be the first discovery that emission lines 
(not just their broad components) could completely disappear from CLAGNs during 
type transition. To verify this interpretation, repeated imaging and better 
spectroscopy will also be needed.

    Lastly, we discuss whether the variabilities could be due to nuclear
transients (e.g., supernovae, tidal disruption events) instead of AGNs. 
First of all, this cannot be the case for \texttt{C3DvarS\_0886}, whose light 
curves show a ``bright-dim-brighter'' behavior. \texttt{C3DvarS\_0722} has a 
similar behavior in F356W, which is the only band that has reliable detections over 
more than two epochs.

    A nuclear transient is also unlikely for \texttt{C3DvarN\_0168}, which 
showed a slow and steady decrease of brightness in F356W and F444W over 723 days 
(rest frame 380.5 days). No known transients would have such a behavior. The 
same reasoning applies to \texttt{C3DvarS\_0723}, whose light curves in the LW
channels collectively show a slow and steady brightness increase over three 
epochs (PRM.2, WEB.2 and C3D.2) spanning 699 days ($\sim$184 days in rest frame).
This is probably also applicable to \texttt{C3DvarN\_0173}. If we consider
its light curves in F150W, F200W, F356W, and F444W collectively, there is also
a slow and steady trend of brightness increase over four epochs spanning
713 days (rest frame 152 days). 

    This leaves \texttt{C3DvarS\_0007} as the only object that we cannot yet rule 
out the possibility of a nuclear transient, but this is because it has reliable
photometry in only two epochs (PRM.2 and C3D.2). Using the F356W band as the
benchmark, the hypothetical nuclear transient appeared in C3D.2 would have 
$m_{356}=28.04$~mag, which corresponds to $M_J\approx -16.84$~mag and falls in
the regime of core-collapse supernovae (CCSNe).

\section{Conclusion and Summary}

   In this work, we conducted a deep near-infrared variability study using the 
public, multi-epoch JWST NIRCam images in the COSMOS field. The imaging data are 
from three programs spanning $\sim$2~years, namely, PRIMER, COSMOS-Web, and 
COSMOS-3D. Within the $\sim$140~arcmin$^2$ of overlapping area, we selected
galaxies that varied by $\geq4\sigma$ in F356W between the PRIMER and COSMOS-3D 
epochs and found 117 variables. Five of them have JWST NIRSpec PRISM 
spectroscopic data from the CAPERS program, which were taken very close in time
to the COSMOS-3D imaging observations (ranging from on the same day to $\sim$153 
days apart). One other has archival spectroscopic data from the 10K-DEIMOS 
survey at the Keck telescope. These six variables, covering a wide redshift 
range of $z\simeq0.2-3.7$ (with the F356W band sampling their 
rest-frame wavelengths $\sim0.76-2.97$~$\mu$m), 
form the spectroscopic sample that is the focus of
this study. Interestingly, two of these six are in a close-pair environment
where the other member did not vary, and two other variables form a close pair 
themselves.

   Variabilities seen in galaxies are usually attributed to AGN. According to the
unification model, type~1 AGNs can vary but type~2 AGNs should not, because our 
view to a type~2 AGN's BLR, which also gives rise to the variable continuum, is 
blocked by its dust torus. This study, while suffering from a few weaknesses due 
to the low spectral resolution, presents a more complicated picture. Among the
six variables in our sample, two are AGNs but only one of them has definite 
evidence for a type~1 (broad Brackett and Pfund series emission lines). 
The other, which is the one that has spectroscopy only 
from 10K-DEIMOS, does not have permitted lines detected. As its host is an 
edge-on disk galaxy, it is most likely a type~2 because type~1 AGNs are rarely 
seen in edge-on disk systems. Two other objects have He~\Romannum{1}$\lambda$10830 
and/or H$\alpha$ detected, and these lines are consistent with being narrow lines.
%While the
%low-resolution of the PRISM data does not allow accurate line width measurements,
%these lines are consistent with being narrow lines, i.e., they could be type~2
%as well. 
A caveat is that their low-S/N spectra could prevent a weak broad-line 
component from being detected (see Appendix~\ref{appendix:linesim}), and
therefore we cannot yet rule out the possibility that they could be type~1.
Nevertheless, the current data support that they could be type~2 as well. 
Finally, there are two objects that have no emission lines at all.
Their light curve behaviors rule out nuclear transients as a possible cause of
their variabilities, and AGN is probably still a viable explanation. 

%{\bf In short, five of the six variables in our sample do not show clear evidence 
%of being type~1 AGNs, 
%which, if confirmed, will add evidence to challenge the AGN unification model. 
%}
    In short, only one of the six variables in our sample is a definite classic
type~1 AGN. Among the other five objects, three are more in line with being type~2 
AGNs, and two do not have any emission lines -- they add evidence to challenge the AGN unification model. 
%{\bf However, as demonstrated in considering the low S/N of the PRISM spectra, 
%broad components can be completely buried in the noise, so the absence of their 
%detections does not rule out the presence of a broad emission line component. 
%}
CLAGNs could be a possible explanation, and the large fraction of non-type-1 
objects in our sample would support that variability search could be an efficient 
method to find candidate CLAGNs. However, the CLAGN scenario would need to explain 
the complete disappearance of emission lines in two of our objects. On the other 
hand, if the two no-line variables are similar to WLQs, they would represent
a new subtype that has low luminosities. With the currently available data, we 
are not able to draw definite conclusions. Regardless, our work shows that 
variability study with JWST is a powerful tool that could lead to new 
understanding of AGNs, and long-term monitoring as well as high-resolution 
spectroscopy will be critical in the future investigations.

\begin{acknowledgements}
    We thank the anonymous referee for useful comments. We also thank 
    Dr. Aigen Li for the discussions on dust in AGNs.
    BS and HY acknowledge the support from the University of Missouri Research Council grant URC-23-029, the NASA grant 80NSSC23K0491, and the NSF grant AST-2307447. 
    
    This project is based on the observations made with the NASA/ESA/CSA James Webb Space Telescope 
    and obtained from the Mikulski Archive for Space Telescopes 
    at the Space Telescope Science Institute, which is operated by the Association of Universities for Research in Astronomy, Inc., under NASA contract NAS 5-03127 for JWST. These observations are associated with program \#5893. Support for program \#5893 was provided by NASA through a grant from the Space Telescope Science Institute, which is operated by the Association of Universities for Research in Astronomy, Inc., under NASA contract NAS 5-03127. We acknowledge the strong support provided by the program coordinator Weston Eck and instrument reviewers Norbert Pirzkal and Stephanie La Massa.
    %which is a collaboration between the Space Telescope Science Institute (STScI/NASA), 
    %the Space Telescope European Coordinating Facility (ST-ECF/ESA), 
    %and the Canadian Astronomy Data Centre (CADC/NRC/CSA). 
    All the JWST NIRSpec data used in this paper can be found in MAST: 
    \dataset[10.17909/26re-7589]{http://dx.doi.org/10.17909/26re-7589}.
\end{acknowledgements}

\appendix

\counterwithin{figure}{section}
\counterwithin{table}{section}

\section{Two Transients Found with Host Spectroscopic Identifications} \label{appendix:transients}

   Our search also revealed 13 transients, two of which have spectroscopic 
identifications of their hosts. We briefly summarize below the behaviors and 
redshift determinations of these two transients. Their coordinates, redshifts, 
and photometry from the difference images are listed in 
Table~\ref{tab:transients}. 

\begin{figure*}[hbt!]
    \centering
    \includegraphics[width=0.9\textwidth,height=0.9\textheight,keepaspectratio]{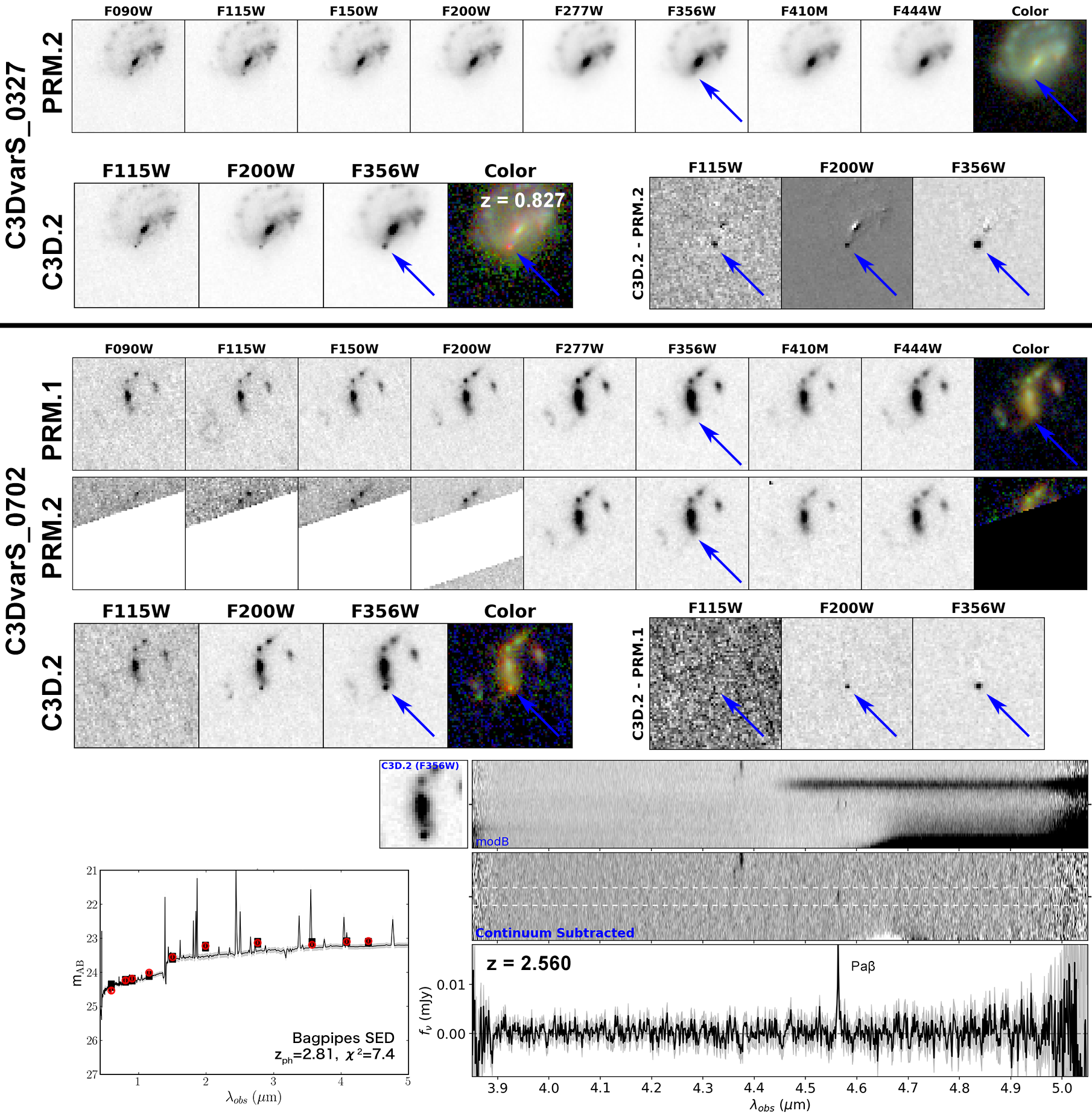}
    \caption{Two transients found in this work that have spectroscopic 
    identifications of their host galaxies. 
    Upper panels: C3DvarS\_0327, a transient in a spiral galaxy at 
    $z_{\rm spec}=0.827$ identified from 3D-HST and zCOSMOS. The image stamps
    on top and bottom are from the reference epoch (PRM.2)
    and the discovery epoch (C3D.2), respectively, and are 3\farcs6$\times$3\farcs6 in size. The difference images
    between the two epochs are also shown in the bottom row on the right. The blue 
    arrow indicates the location of the transient.
    Lower panels: C3DvarS\_0702, a transient within a potential 
    group of galaxies. The image stamps on the top three rows are similar to
    the above. The C3D WFSS 2D spectrum, before and after the continuum 
    subtraction, is also shown. A single line is revealed, which we identify
    as $\rm Pa\beta$ based on the consistency with the SED fitting result as 
    shown in the bottom-left. This gives $z_{\rm spec}=2.560$, which is labeled 
    in the 1D spectrum at the bottom-right.
    }
    \label{fig:transients}
\end{figure*}

$\bullet$ {\tt C3DvarS\_0327}: this transient was invisible in the PRM.2 epoch 
and appeared in the C3D.2 epoch. It is located in a highly asymmetric spiral 
galaxy at $z_{\rm spec}=0.827$. The spectroscopic redshift was independently 
measured and confirmed by both the 3D-HST 
\citep[][]{Brammer2012_3dhst,Momcheva2016_3dhst} 
and zCOSMOS \citep[][]{Lilly2007_zcosmos,Lilly2009_zcosmos-brt} surveys. At
this redshift, its F115W magnitude corresponds to $M_R\approx -16.4$~mag, which
falls in the regime of CCSNe. It was very red in the C3D.2 epoch, with 
$m_{115}-m_{356}=1.56\pm0.09$~mag. The upper panels of 
Figure~\ref{fig:transients} show their image stamps in the two epochs as well 
the difference images in the three common bands. Note that it is very close a 
faint knot of the host, which was not varying.

$\bullet$ {\tt C3DvarS\_0702}: this object is shown in the lower panels of
Figure~\ref{fig:transients}. It occurred on the outskirts of a compact galaxy 
that is likely associated with a group of smaller galaxies extending to its 
northwest. The transient was not detected in the PRM.1 and PRM.2 epochs and 
appeared in the C3D.2 epoch. As it fell in the detector gap in all SW channels 
in the PRM.2 epoch, the difference images shown in Figure~\ref{fig:transients}
are constructed between C3D.2 and PRM.1. The C3D NIRCam WFSS Grism R spectrum 
(in F444W) revealed a single emission line at 
$\lambda_{\rm obs}=4.564\mu{\rm m}$, but this line alone is insufficient for 
redshift determination. To find its most likely redshift, we fitted the 
SED of the host galaxy using 
Bagpipes \citep[version 1.2.0; ][]{Carnall18_bagpipes}, which yielded
$z_{\rm phot}=2.81$ with a small $\chi^2=7.4$. 
Therefore, we interpreted the single emission line as $\rm Pa\beta$, 
which thus gives $z_{\rm spec}=2.560\pm0.002$ and
$|\Delta z|/(1+z_{\rm spec})=0.07$. At this redshift, its F150W magnitude
corresponds to $M_B \approx -18.2$~mag, which is likely a Type~Ia supernova. It
also had a red color in the discovery epoch, with 
$m_{150}-m_{356}=0.92\pm0.12$~mag.

\begin{table*}[hbt!]
    \raggedright
    \caption{Positions, redshifts and photometry of two transients}
    \begin{tabular}{lcccccc} \hline 
        SID & R.A. & Decl. & $z_{\rm spec}$ & $m_{115}$ & $m_{150}$ & $m_{356}$ \\ \hline 
        C3DvarS\_0327 & 150.1215211 & 2.2670583 & 0.827 & $26.57\pm0.09$ & $26.00\pm0.05$ & $25.01\pm0.02$ \\
        C3DvarS\_0702 & 150.1377801 & 2.3608334 & $2.560\pm0.002$ & $>27.70$ & $27.03\pm0.10$ & $26.11\pm0.06$ \\
        \hline 
    \end{tabular}
    \tablecomments{
    Spectroscopic redshifts are based on the host galaxies. Photometry for 
    \texttt{C3DvarS\_0327} was obtained using the C3D.2$-$PRM.2 difference 
    images. \texttt{C3DvarS\_0702} fell in the detector gap in the PRM.2 epoch 
    in all SW bands, and therefore its photometry was done on the C3D.2$-$PRM.1 
    difference images. We also checked the measurement in F356W on the 
    C3D.2$-$PRM.2 difference image for the latter, which gave 
    $m_{356}=26.16\pm0.05$, consistent with the measurement presented in this table. 
    }
    \label{tab:transients}
\end{table*}

\section{Resolving Power of the NIRSpec PRISM Mode}\label{appendix:pr}

   During our line width measurements, we found many cases where the obtained
FWHM values are significantly narrower than the predictions given by the NIRSpec 
PRISM resolving power curve available at the JDOC site.
As this curve is based on the pre-launch estimates, our results indicate that
the in-flight performance of the PRISM mode is significantly better. Our
consultation with the JWST Help Desk confirmed that this was indeed the case. 
Basically, this is because the JWST has achieved better (i.e., smaller) point 
spread functions than the pre-launch expectations. 
This makes the spectral resolution of NIRSpec better than the prediction across
all wavelengths, and the improvement is the most obvious in the PRISM mode due
to its low resolution.

    Unfortunately, the accurate in-flight measurements of the NIRSpec resolving 
powers are not yet available as of this writing. To better constrain the 
intrinsic line widths needed for this study, we had to estimate the actual PRISM 
resolving powers on our own using the archival data. The idea was to use the 
same narrow lines that have observations in both the PRISM mode and the 
medium-resolution grating mode. As the resolving powers of the two differ by 
$\sim$10$\times$ ($R\approx 30$--100 versus $R\approx 1000$), it is a good 
approximation to take the line width measured in the grating mode 
(${\rm FWHM_{og}}$) as the true line width (${\rm FWHM_t}$). Therefore, the 
PRISM line broadening factor (${\rm FWHM_{pb}}=\lambda/R$) can be obtained 
through the observed line width in the PRISM mode (${\rm FWHM_{op}}$) by using 

\begin{align}
    {\rm FWHM_{pb}=\sqrt{\rm FWHM_{op}^2-FWHM_{og}^2}}.
\end{align}

    To this end, we utilized the emission line galaxy sample compiled in 
\citet[][]{SY2025b} to select lines suitable for this purpose. We carefully 
chose 31 galaxies at different redshifts to cover the wavelength range spanned 
by our variable sample. We used mostly $\rm H\alpha$ because it is the strongest 
among single lines. We visually inspected the grating spectra (in G140M, G235M 
or G395M) to ensure that no nearby lines (e.g., [N~\Romannum{2}]) could 
contaminate the line width measurement in the PRISM spectrum. To cover the 
wavelengths bluer than 1.5~$\mu$m, we had to use the [O~\Romannum{2}] line. While
it is a doublet, the two lines are only 2.8\AA\ apart in the rest frame and
are effectively a single line in the grating spectra.

\begin{figure}
    \centering
    \includegraphics[width=0.8\linewidth]{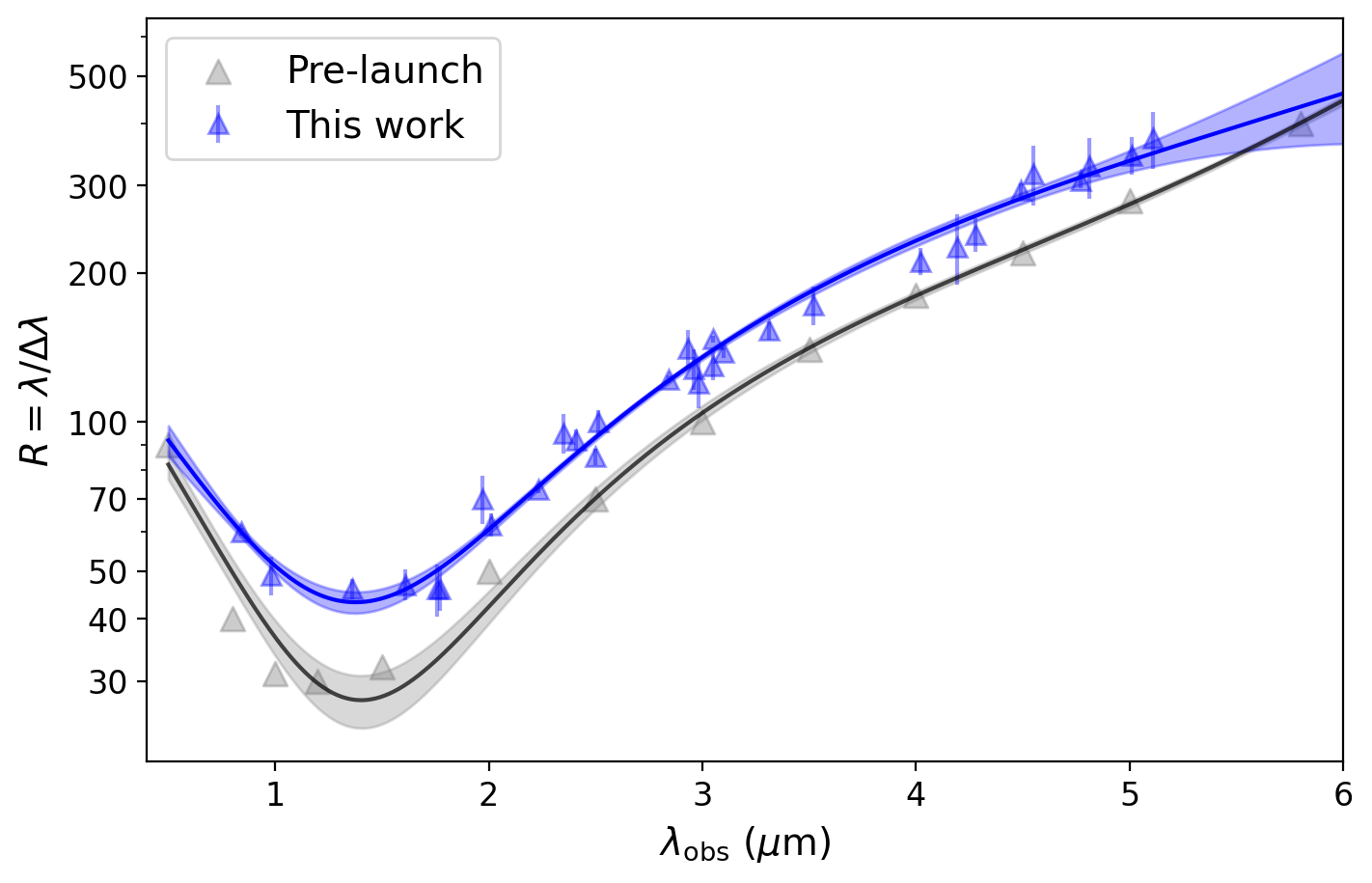}
    \caption{Comparison between the pre-launch PRISM resolving power (gray triangles, retrieved from the JDOC site) 
    and our estimates (blue triangles). 
    The gray and blue curves represent the best-fit fourth-order polynomial fits 
    and the shaded areas show the $1\sigma$ uncertainty. 
    }
    \label{fig:resolve-power}
\end{figure}

    Figure~\ref{fig:resolve-power} shows our estimates of the resolving power
in blue triangles. For comparison, we digitized the JDOC pre-launch resolving 
power curve at 13 representative wavelengths, which are shown as the gray 
triangles. These gray triangles can be well fitted by a fourth-order polynomial 
(shown as the black curve), and therefore, we fitted the same polynomial to the 
blue triangles and obtained the blue curve. This curve allowed us to estimate 
the PRISM resolving power (and therefore the line broadening in Section 3.2.2) 
at any wavelength. Note that in reality ${\rm FWHM_{og}} > {\rm FWHM_t}$, and
therefore ${\rm FWHM_{pb}}$ is slightly underestimated and $R$ is slightly 
overestimated. 

\section{SEDs in Luminosity Units and Difference Image Light Curves}
\label{appendix:altplot}

   Here we show the SEDs of the six objects in monochromatic luminosities 
($\nu L_\nu$) and their ``light curves'' measured in difference images. The latter 
reflect the net amplitudes of the variabilities.

\begin{figure}[hbt!]
    \centering
    \includegraphics[width=0.9\linewidth]{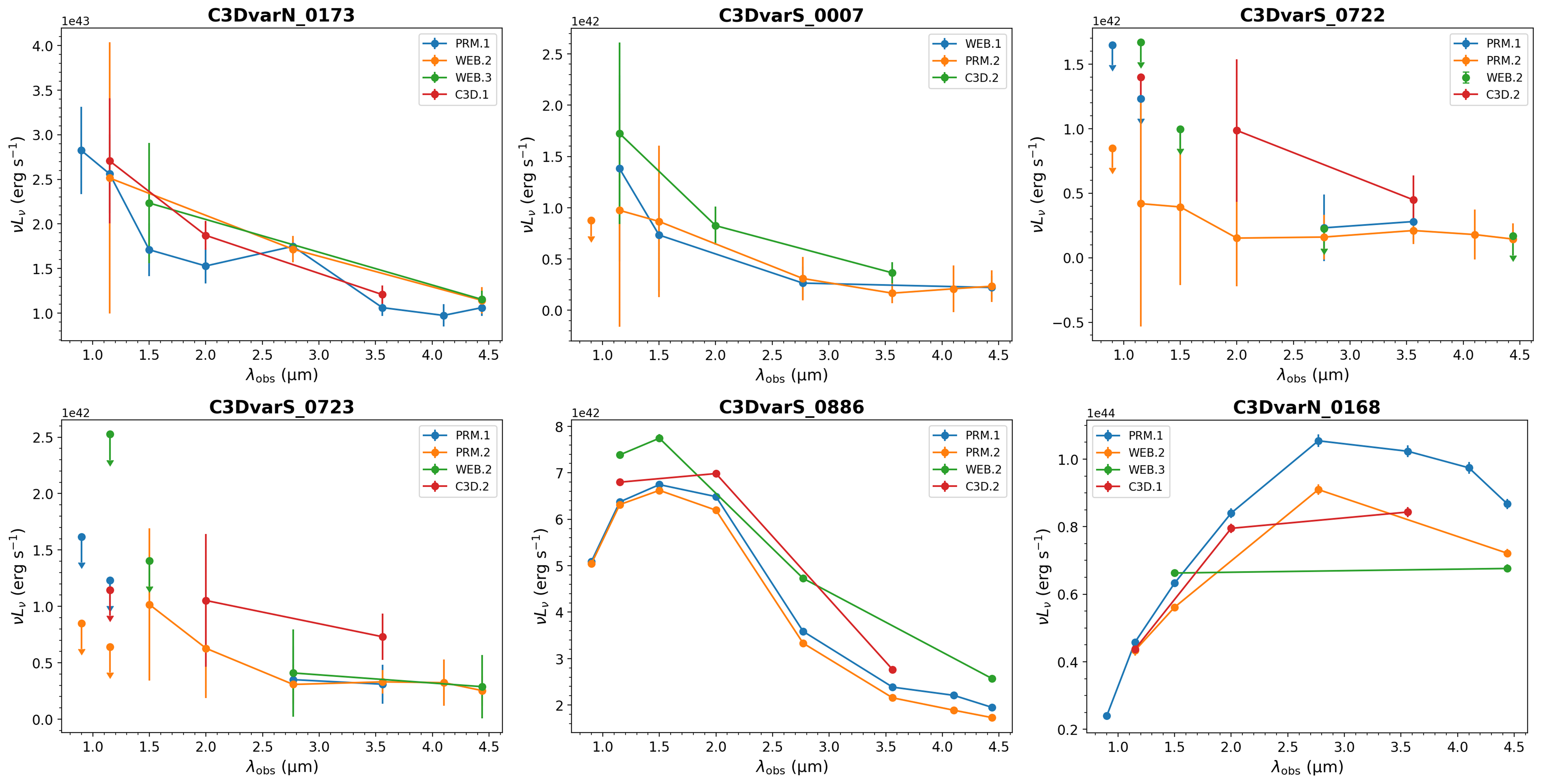}
    \caption{Spectral energy distributions of the six variables in luminosity units ($\nu L_\nu$). 
     Data points with downward arrows indicate the $2\sigma$ upper limits. 
    }
    \label{fig:nuLnu_SED}
\end{figure}

   Firstly, for each source and in each epoch, we converted the observed 
$m_{\rm AB}$ into monochromatic luminosities $\nu L_\nu$, and the results are shown
in Figure~\ref{fig:nuLnu_SED}. 

\begin{figure}[hbt!]
    \centering
    \includegraphics[width=0.9\linewidth]{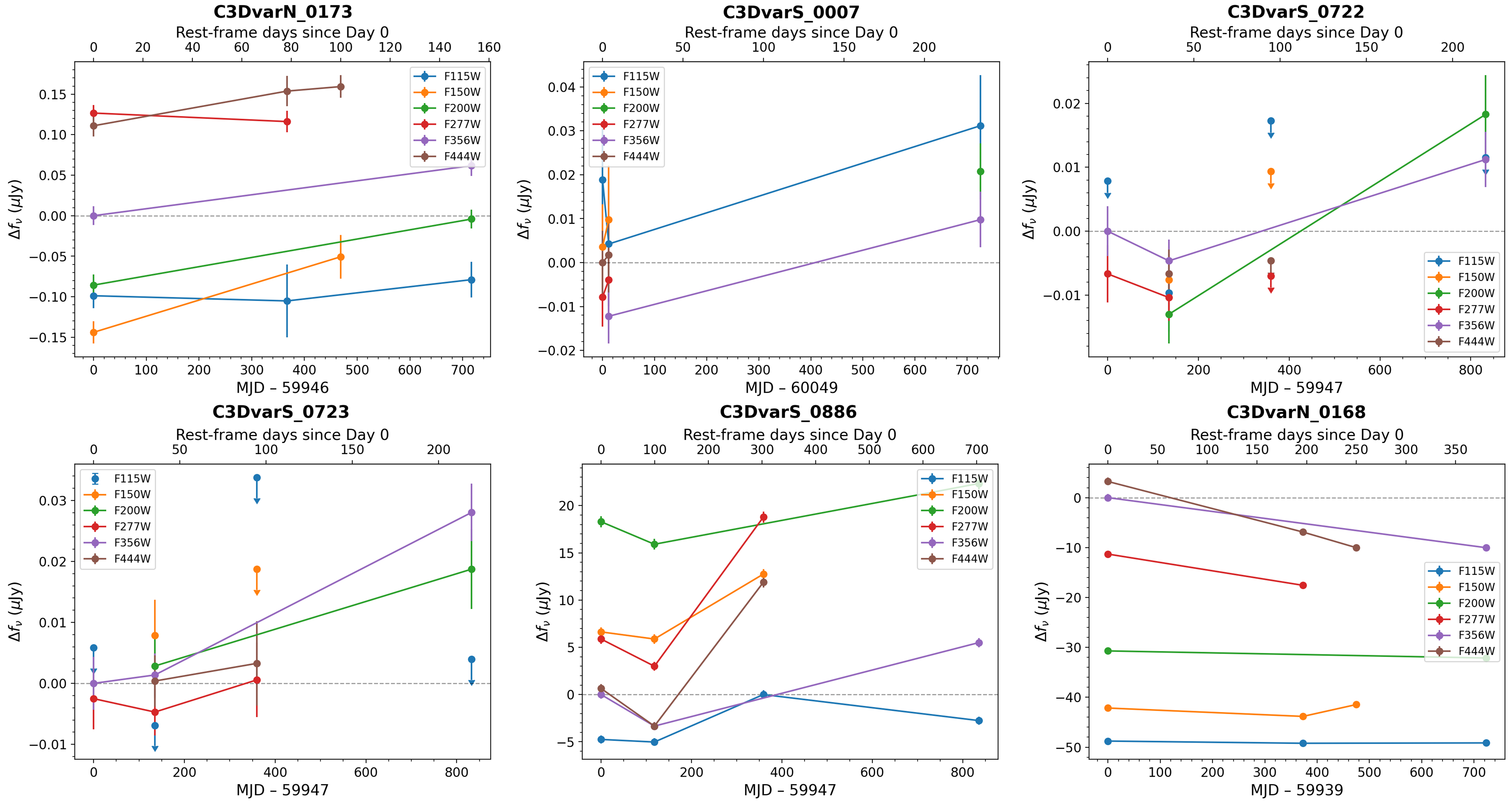}
    \caption{
    Light curves measured from difference images. For each source, 
    we plotted the change of flux density $\Delta f_\nu$ relative to a reference 
    (F356W or F444W photometry in the first epoch). 
    The rest-frame times are shown on the top axis. 
    In the cases where the source is undetected in one of the epochs 
    and thus only an upper limit (downward arrows) on the flux change can be estimated. 
    }
    \label{fig:delta_flux}
\end{figure}

    We also obtain ``light curves'' by doing photometry in difference images. 
For each variable, we first constructed a segmentation map with {\sc photutils} 
using the F356W difference image between its detection and reference epochs. 
We then used this segmentation map to all other difference images in all bands and 
epochs so that the aperture is consistent. The results are shown in 
Figure~\ref{fig:delta_flux}. For each source, we adopt the F356W photometry in the 
first epoch as the reference ($\Delta f_\nu=0$); for C3DvarS\_0007 where the first epoch does not have F356W, we used F444W as the reference. 

\section{Possibility of a Weak Broad Component Undetected in Noisy Data}\label{appendix:linesim}

   We carried out a simple simulation to examine whether a broad emission 
line component could be present in \texttt{C3DvarN\_0173} and 
\texttt{C3DvarS\_0007} but is undetected due to the low S/N of their spectra.

    The simulated line profile is the sum of two Gaussian profiles: a narrow 
component with $\Delta v$ measured from the spectrum (as listed in 
Table~\ref{tab:line_measure}) and a broad component with $\Delta v$ varying 
from 1000 to 10000~km~s$^{-1}$. The noise level of the narrow component was 
so chosen that the line has the same S/N as the real measurement in 
Table~\ref{tab:line_measure}. For each choice of the broad line width, 
we assigned the broad component an integrated flux that is $1/3$ of the 
narrow line. 
     
\begin{figure}[hbt!]
    \centering
    \includegraphics[width=0.3\linewidth]{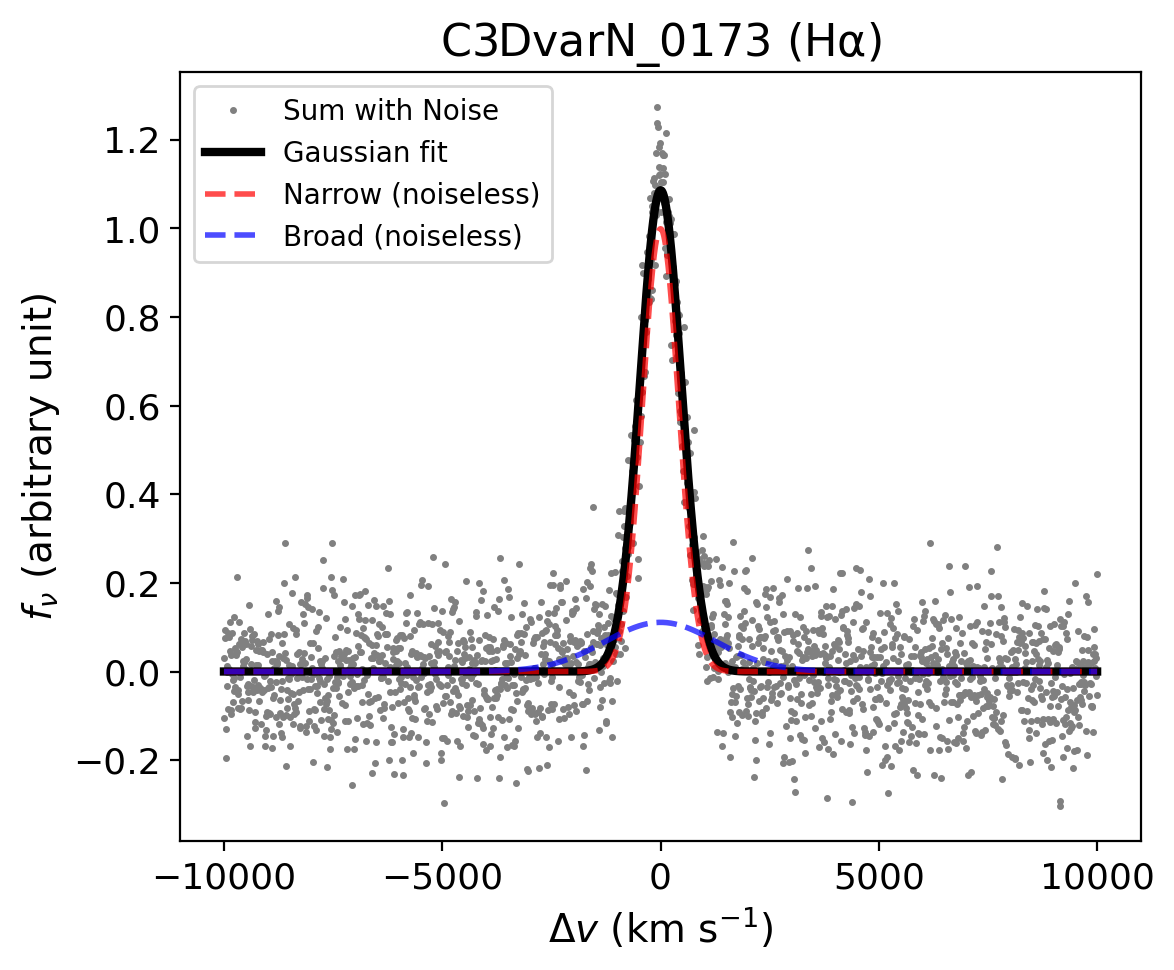}
    \includegraphics[width=0.3\linewidth]{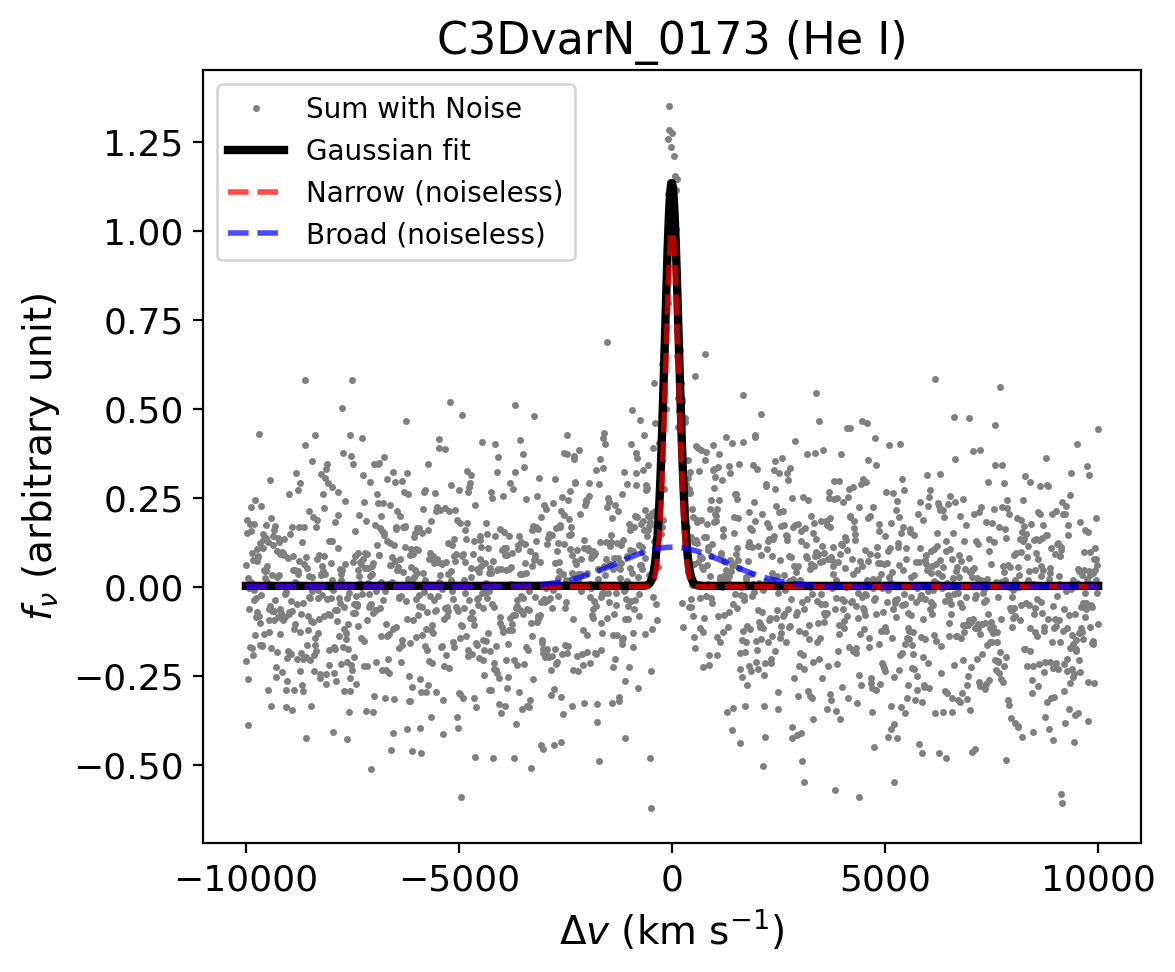}
    \includegraphics[width=0.3\linewidth]{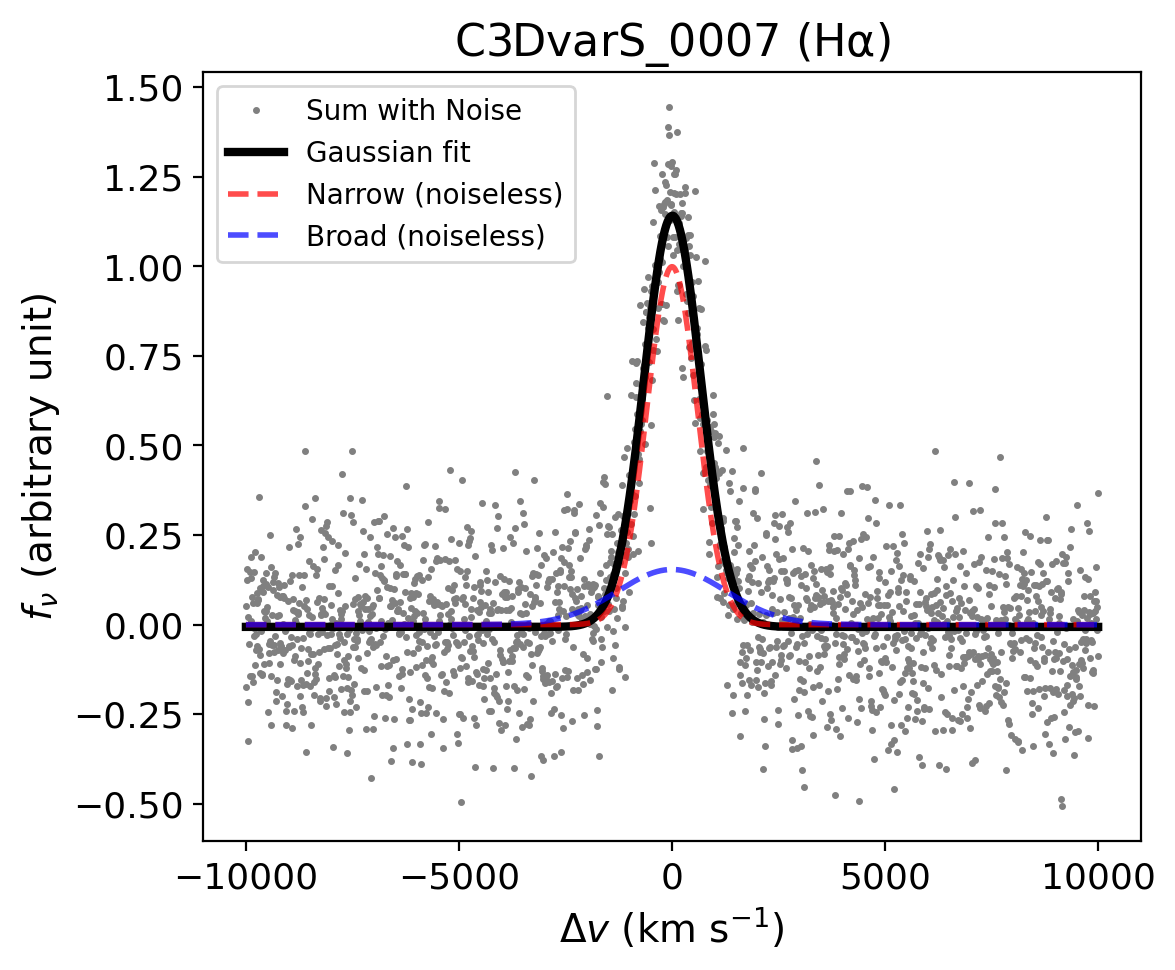}
    \caption{
    Simulated emission line profiles for \texttt{C3DvarN\_0173} and
    \texttt{C3DvarS\_0007} and single Gaussian fits to these 
    profiles. Each line is modeled as the sum of a narrow Gaussian 
    ($\Delta v$ as measured in Table~\ref{tab:line_measure}) and a broad 
    Gaussian with $1/3$ integrated flux of the narrow component and 
    $\Delta v=3000$~km~s$^{-1}$ for demonstration purpose ($\Delta v$ ranging 
    from 1000 to 10000~~km~s$^{-1}$ in the actual simulation). In each panel,
    the gray data points show the summed profile with noise added, the red 
    and blue dashed curves show the noiseless narrow and broad components,
    respectively, and the black solid curve shows the best-fit single 
    Gaussian profile to the simulated line. 
    }
    \label{fig:linesim}
\end{figure}

    Figure~\ref{fig:linesim} demonstrates the results of this simulation for
the three permitted lines in the case of broad line width 
$\Delta v=3000$~km~s$^{-1}$. Each panel shows the simulated line with noise 
added (gray points), the noiseless narrow and broad components (red and blue 
dashed curves, respectively), and the single Gaussian profile fit (black 
curve) to the simulated line. Clearly, the fit yields a narrow-line-like 
width with no obvious evidence of a broad component. This is true even when 
we set the width of the broad component to $\Delta v=1000$~km~s$^{-1}$. 
We also attempted to fit the simulated line with a double-Gaussian model, 
however, the best-fit parameters for the broad component are highly
unreliable (with errors comparable or larger than the fitted values), which 
means that the broad component is not constrained. In summary, the broad 
component cannot be discerned in the presence of noise as in our current data.
Therefore, we cannot rule out the possibility of an underlying broad 
component in the permitted lines of C3DvarN\_0173 and C3DvarS\_0007. 

\section{Stellar and AGN Components}
We also attempt to investigate whether the variable objects could have an AGN
component by fitting their SEDs. For this purpose, we employed {\sc CIGALE} 
\citep[version 2022.1;][]{Boquien19_cigale}, which has a functionality to
separate stellar and AGN contributions when fitting SEDs. We use the 
photometry in the PRM epochs to construct the SEDs (the other two epochs only
have a fraction of the passbands in the PRM epochs). The galaxy templates are 
generated using the population synthesis models of
\citet[][BC03]{Bruzual2003} with the initial mass function of 
\citet[][]{Chabrier2003} and a delayed-$\tau$ star formation history following
SFR~$\propto t e^{-t/\tau}$ with $0.01\leq \tau \leq 13$~Gyr.
The redshift is set to the spectroscopic redshift in Table~\ref{tab:var_info}.
To avoid overfitting, we fix the metallicity to $Z=0.02$ and set the nebular 
emission contribution with a fixed $\log(U)=-2.5$. We adopt the modified 
Calzetti extinction law in CIGALE's {\tt dust\_modified\_starburst} module, 
with the nebular gas reddening $E(B-V)_g$ ranging from 0 to 3~mag at a step 
of 0.2~mag and a fixed multiplication factor of 0.44 to calculate the stellar 
continuum attenuation $E(B-V)_s=0.44\times E(B-V)_g$. 
The AGN templates are the {\tt skirtor2016} models from 
\citet[][]{Stalevski12_agn,Stalevski16_agn}, 
and we vary the AGN fraction ($f_{\rm AGN}$) from 0 to 1 at a step size of 
0.2. The viewing angles $i$ are set to 30$^\circ$ and 70$^\circ$ for Type 1 
and 2 AGNs, respectively. 
%We noticed that the program runs into errors when $f_{\rm AGN}=1$, 
%and we resolve this issue by setting the maximum $f_{\rm AGN}$ to 0.999. 

\begin{figure}
    \centering
    \includegraphics[width=0.9\linewidth]{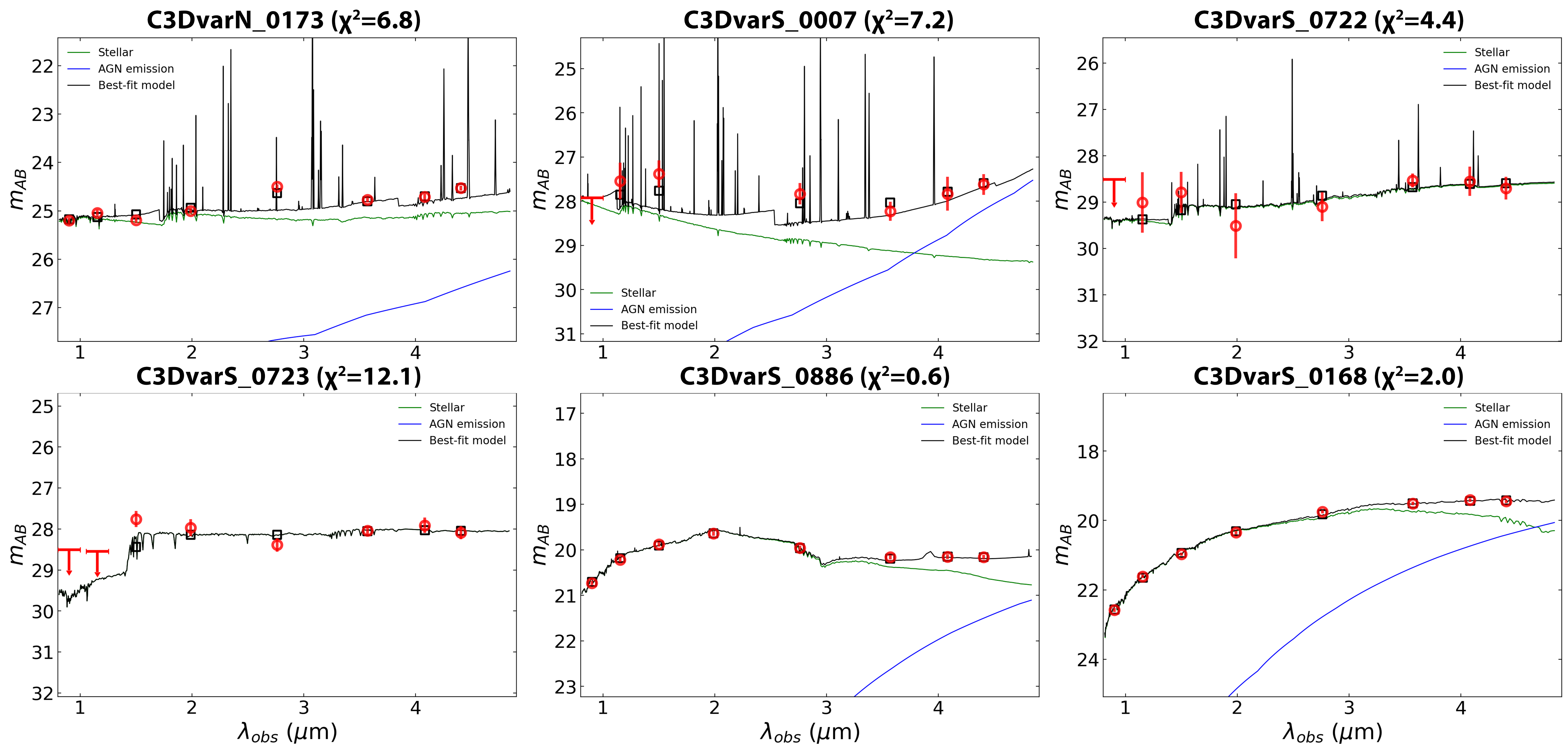}
    \caption{{\sc CIGALE} SED fitting results on the host galaxies of the six 
    variables in our spectroscopic sample. The red symbols represent the 
    observed magnitudes, and the black curves are the best-fit models. The 
    black rectangles are the synthetic magnitudes derived from the 
    corresponding best-fit models. 
    }
    \label{fig:sedfit}
\end{figure}

    The SED fitting results are shown in Figure~\ref{fig:sedfit},
and the best-fit parameters are listed in Table~\ref{tab:sedfit}. 
Note that CIGALE's $f_{\rm AGN}$ is defined as the fractional contribution
of the AGN component to the total IR luminosity $L_{\rm IR}$
(rest-frame 8--1000~$\mu$m). Therefore, it could happen that an object with a
high $f_{\rm AGN}$ has only a minimal AGN contribution in the rest-frame
optical-to-near-IR regime. To obtain the AGN fraction in the wavelength
range covered by our SEDs, we calculate 
$f_{\rm AGN,nircam}=L_{\rm AGN,nircam}/L_{\rm tot,nicam}$, where 
$L_{\rm AGN,nircam}$ and $L_{\rm tot,nicam}$ are the luminosity of the AGN
component and the total luminosity, respectively, obtained by integrating the
best-fit AGN template and the total template over the NIRCam wavelength 
range, respectively. This value is also listed in the table.

    As the SED wavelength coverage and the number of passbands are limited,
these fitted parameters should be taken with caution. For instance, the
age and SFR values for the stellar components all have large errors. As 
another example, \texttt{C3DvarS\_0722/0723} have negligible AGN 
contributions to the NIRCam SEDs ($f_{\rm AGN,nircam}$) despite the high 
$f_{\rm AGN}$.

   Assuming that the decomposition of the stellar and AGN components 
mentioned above is reasonable, we attempt to estimate their black hole masses 
($M_{\rm BH}$) by using the scaling relation between $M_{\rm BH}$ and the 
host galaxy's stellar mass ($M_*$) of 
\citet[][their Eqn. 4 \& 5]{ReinesVolonteri2015}. These values are listed
in the last column of the table.

\begin{table}[hbt!]
    \centering
    \scriptsize
    \begin{tabular}{lccccccccc}
        SID & $\log_{10}(M_*/M_\odot)$ & $Age$ (Myr) & SFR ($M_\odot$/yr) & E(B-V) & $f_{\rm AGN}$ & $f_{\rm AGN,nircam}$ & $i$ ($^{\rm o}$) & $\log_{10}(M_{\rm BH}/M_\odot$) \\ \hline 
        C3DvarN\_0173 & $8.85\pm0.15$ & $28\pm80$ & $87.54\pm49.59$ & $0.20\pm0.05$ & $0.81\pm0.09$ & $0.09$ & $74\pm13$ & $5.19\pm0.16$ \\ 
        C3DvarS\_0007 & $7.52\pm0.40$ & $482\pm709$ & $0.63\pm0.71$ & $0.10\pm0.08$ & $0.82\pm0.10$ & $0.27$ & $30\pm25$ & $3.80\pm0.42$ \\ 
        C3DvarS\_0722 & $7.69\pm0.38$ & $580\pm626$ & $0.51\pm0.82$ & $0.22\pm0.16$ & $0.81\pm0.09$ & $<0.01$ & $29\pm25$ & $3.97\pm0.40$ \\ 
        C3DvarS\_0723 & $7.92\pm0.16$ & $669\pm597$ & $0.22\pm0.36$ & $0.05\pm0.07$ & $0.83\pm0.10$ & $<0.01$ & $31\pm25$ & $4.22\pm0.17$ \\ 
        C3DvarS\_0886 & $9.69\pm0.12$ & $5929\pm2663$ & $0.01\pm0.02$ & $0.16\pm0.05$ & $0.75\pm0.06$ & $0.09$ & $67\pm22$ & $6.07\pm0.13$ \\ 
        C3DvarN\_0168 & $10.85\pm0.12$ & $2656\pm1603$ & $22.09\pm28.64$ & $0.53\pm0.12$ & $0.79\pm0.08$ & $0.23$ & $68\pm16$ & $7.29\pm0.13$ \\ 
        \hline 
    \end{tabular}
    \caption{Stellar and AGN component decomposition results corresponding
    to the {\sc CIGALE} SED fitting shown in Figure~\ref{fig:sedfit}. The
    derived parameters are based the Bayesian estimates directly returned from {\sc CIGALE}. 
    }
    \label{tab:sedfit}
\end{table}

%\begin{table}[hbt!]
%    \centering
%    \begin{tabular}{lcccccccc}
%         SID & Line & $I_{\rm narrow}/3$ & $I_{\rm narrow}/5$ & $I_{\rm narrow}/10$ \\ %\hline 
%         C3DvarN\_0173 & $\rm H\alpha$ & $2450$ & $<1000$ & $<1000$ \\ 
%          & He I & $<1000$ & $<1000$ & $<1000$ \\ 
%          \hline 
%         C3DvarS\_0007 & $\rm H\alpha$ & $9486$ & $3359$ & $<1000$ \\ 
%          \hline 
%    \end{tabular}
%    \caption{Maximum line width that can be measured with S/N~$=3$ 
%     assuming the broad component of the emission line has a flux at $1/3$, $1/5$, and %$1/10$ 
%     the intensity of the flux of the narrow component (see Table~\ref{tab:line_measure}). 
%    }
%    \label{tab:placeholder}
%\end{table}

\bibliography{sample631}{}
\bibliographystyle{aasjournal}

\end{document}